\documentclass[pra,twocolumn,preprintnumbers,amsmath,amssymb]{revtex4}

\usepackage{graphicx}% Include figure files
\usepackage{dcolumn}% Align table columns on decimal point
\usepackage{bm}% bold math
\usepackage{color}
\usepackage{epstopdf}
\usepackage{float}
\usepackage{ulem}
\usepackage{lipsum,babel}
\newcommand{\blue}{}

\begin{document}

%\widetext

%\preprint{APS/123-QED}

\title{Irreversible thermalization vs reversible dynamics mediated by anomalous correlators: Wave turbulence theory and experiments in optical fibers}
\author{T. Torres$^{1}$, J. Garnier$^{2}$, L. Zanaglia$^{3}$, M. Ferraro$^{1,4}$, C. Michel$^{3}$, V. Doya$^{3}$,\\ J. Fatome$^{1}$, B. Kibler$^{1}$, S. Wabnitz$^{5}$, A. Picozzi$^{1}$, G. Millot$^{1}$}
\affiliation{$^{1}$ Universit\'e Bourgogne Europe, CNRS, Laboratoire Interdisciplinaire Carnot de Bourgogne ICB UMR 6303, 21000 Dijon, France}
\affiliation{$^{2}$CMAP, CNRS, Ecole polytechnique, Institut Polytechnique de Paris, 91120 Palaiseau, France}
\affiliation{$^{3}$Universit\'e C\^ote d'Azur, CNRS, Institut de Physique de Nice, Nice, France}
\affiliation{$^{4}$Department of Physics, University of Calabria, Via P. Bucci, Rende, 87036, (CS), Italy}
\affiliation{$^{5}$Department of Information Engineering, Electronics, and Telecommunications, Sapienza University of Rome, Via Eudossiana 18, Rome, 00184, Italy}
%\affiliation{$^{3}$Institut Universitaire de France (IUF), 1 rue Descartes, 75005 Paris, France}
%\affiliation{$^{6}$INO-CNR Pitaevskii BEC Center and Dipartimento di Fisica, Universit\`a di Trento, 38123 Povo, Italy}
%\affiliation{$^{1}$ Universit\'e Bourgogne Europe, CNRS, Laboratoire Interdisciplinaire Carnot de Bourgogne ICB UMR 6303, F-21000 Dijon, France}
%\affiliation{$^{2}$ Universit\'e C\^ote d'Azur, CNRS, Institut de Physique de Nice, Nice, France}
%\affiliation{$^{3}$ CMAP, CNRS, Ecole Polytechnique, Institut Polytechnique de Paris, 91128 Palaiseau Cedex, France}
%\affiliation{$^{4}$ CEA, DAM, DIF, F-91297 Arpajon Cedex, France} 
%\affiliation{$^{5}$ Institute of Physical Chemistry Polish Academy of Sciences, Warsaw, Poland}
%\affiliation{$^{5}$ Institut Universitaire de France (IUF), 1 rue Descartes, 75005 Paris, France}


\begin{abstract}
We theoretically and experimentally investigate spontaneous self-organization in a conservative (Hamiltonian) turbulent wave system, operating far from thermodynamic equilibrium. Our system is governed by two coherently coupled nonlinear Schr\"odinger equations, describing the polarization evolution of light in a dispersive nonlinear optical fiber. The analysis reveals the emergence of two fundamentally distinct turbulent regimes. In a first regime, the waves undergo a slow, irreversible thermalization process, which is accurately described by the wave turbulence kinetic equation and the associated $H-$theorem of entropy growth. In stark contrast with this expected irreversible process, we identify a second different regime, where strong phase-correlations spontaneously emerge, giving rise to a fast reversible oscillatory dynamics of the normal correlator and anomalous phase-correlator. Experimental observations confirm the occurrence of both irreversible thermalization and reversible dynamics mediated by the anomalous correlated fluctuations. 
%In stark contrast, we identify a second different regime, where strong phase-correlations spontaneously emerge, giving rise to a fast reversible oscillatory dynamics of the normal correlator and anomalous phase-correlator. Experimental observations confirm the occurrence of both irreversible thermalization and reversible dynamics mediated by the anomalous correlated fluctuations. 
%These findings shed new light on spontaneous self-organization phenomena in conservative (Hamiltonian) turbulent wave systems operating far from thermodynamic equilibrium.
\end{abstract}

%\pacs{42.65.Sf, 05.45.a}

\maketitle

{\it Introduction.} 
Non-integrable Hamiltonian systems of random waves generally undergo thermalization -- an irreversible evolution toward thermodynamic equilibrium, characterized by a state of maximum entropy. In the weakly nonlinear regime, this process is accurately described by the wave turbulence (WT) theory \cite{zakharov92,zakharov04,nazarenko11,Newell_Rumpf,galtier22,onorato23}, which has been successfully applied to various physical systems \cite{lvov97,rumpf09,nazarenko25,nazarenko17,onorato20,onorato15}, including optical waves \cite{laurie12,babin07,turitsyn13,churkin15,PRL10,PR14}. 
%Despite formal Hamiltonian reversibility, 
The WT kinetic equation (KE) describes the actual irreversible evolution to the Rayleigh-Jeans equilibrium distribution, which is expressed by a $H-$theorem of entropy growth.
Interestingly, light thermalization has recently been studied theoretically \cite{PRL19,christodoulides19,podivilov22} and experimentally \cite{podivilov22,PRL20,pourbeyram22,mangini22} in multimode fibers, thus spurring the emerging field of optical thermodynamics \cite{christodoulides19,PRL23,muniz23,ferraro24,kirsch25,PRL22,lu24,kottos20,kottos23,kottos24,efremidis24,ferraro25_rev}.
%, see the reviews~\cite{wrightNP22,ferraro23,ferraro25_rev}.
%, in analogy with the celebrated Boltzmann's $H-$theorem in kinetic gas theory.
%On the basis of theoretical works \cite{PRL19,podivilov19,PRA19,christodoulides19}, RJ thermalization has recently been experimentally studied %in the 2D spatial dynamics of a speckle beam propagating in multimode optical fibers \cite{PRL20,pourbeyram22,mangini22,podivilov22}, on the basis of the beam cleaning effect \cite{wright16,krupa17} and related 
%thus spurring the emergent key area of optical thermodynamics \cite{christodoulides19,wrightNP22,PRL23,muniz23,ferraro24,kirsch25,PRL22,lu24,kottos24,efremidis24,ferraro23,ferraro25_rev}. 
%In the context of hydrodynamic wave tubulence, experimental evidence of large scale statistical thermal equilibrium has been also recently reported \cite{fauve17,falcon25}.  

Although rigorous derivations of the WT KE have been accomplished for certain partial differential equations (PDEs) \cite{deng23full,deng23long,staffilani21}, the broader validity of the WT theory across a wide range of nonlinear wave-interacting PDEs remains unclear. A central requirement in deriving the KE is the demonstration that {\it phase-correlations} among the random waves can be neglected. 
%To date, this condition has been rigorously verified only for the scalar Nonlinear Schr\"odinger (NLS) equation in dimensions $d \gtrsim 2$ [8].
%Fourier phases of the normal variables stay uncorrelated over kinetic timescales. 

Phase-correlations are known to underlie
%play a key role in 
large-scale coherent structures, such as solitons or condensates, which emerge from strongly nonlinear turbulent regimes \cite{zakharov04,nazarenko11,Newell_Rumpf,galtier22,onorato23,laurie12,lvov97,rumpf09,turitsyn12,Turitsyna13,mordant23,delre16,delre24}. These coherent structures are, by their nature, phase-correlated entities. In contrast, we will focus here on phase-correlations in the 
{\it weakly nonlinear regime} of random dispersive waves.
%In contrast, we focus here on {\it weakly interacting random waves}, where phase-correlations are expected to decay over time (or propagation distance), as commonly assumed in WT theory.

%A special type of phase correlation that can arise is the so-called anomalous correlator. 
%Phase-correlations, also called `anomalous correlations', originally appeared in the BCS theory of superconductivity \cite{bcs57}, and then have been studied in the framework of the $S-$theory \cite{zakharov75,lvov94}, in particular to describe the parametric excitation of magnet systems from an external source \cite{lvov94}. 
Phase-correlations -- also referred to as `anomalous correlations' -- were central to the original Bardeen–Cooper–Schrieffer theory of superconductivity \cite{bcs57}. They have since been extensively studied within the framework of $S-$theory \cite{zakharov71,zakharov75,lvov94}, where anomalous correlations arise from an external forcing (pumping) that parametrically excites a dissipative magnetic system \cite{lvov94}.
\blue{Anomalous correlators connected to optical polarization were originally introduced in Ref.~\cite{kuznetsov78} to study 
turbulence of electromagnetic waves in isotropic plasma.}
%such phase-correlations arise due to an external forcing (pumping) that parametrically excites a dissipative magnetic system \cite{lvov94}.
From a different perspective, recent theoretical advances revealed the presence of phase-correlations in conservative (unforced) wave systems.
%In the vector nonlinear Schr\"odinger equation (NLSE), strong convective coupling between wave components drives these anomalous correlations \cite{guasoni17}. Phase correlations have been also identified in particle chains governed by the %FPUT Fermi-Pasta-Ulam-Tsingou model \cite{zaleski20}. 
Anomalous phase-correlations have been identified in Fermi–Pasta–Ulam–Tsingou chains~\cite{zaleski20}, or in the coupled nonlinear Schr\"odinger equation (NLSE), arising from strong convection between wave components~\cite{guasoni17}.
%Recently, the mechanism underlying the emergence of phase correlations between different frequencies has been elucidated by showing that they arise from Hamiltonian terms that break phase invariance, whereas systems that preserve phase invariance preclude such anomalous correlations~\cite{villois25}.
In {\it scalar} wave systems, the mechanism underlying the emergence of phase-correlations between different frequencies has been recently elucidated, by showing that they may only arise from Hamiltonian terms that break phase invariance, whereas systems that preserve phase invariance preclude such anomalous correlations~\cite{villois25}. 
%Despite substantial progress, the origin of anomalous correlators and their role in the thermalization of conservative wave systems remain unexplored, representing a key open problem in nonequilibrium statistical physics.

In this Letter, we investigate a {\it vector} system of coherently coupled NLSEs, which describe the polarization evolution of temporally incoherent light waves in optical fibers. We identify two fundamentally different regimes: an irreversible thermalization, and a reversible turbulent dynamics driven by phase-correlations.  
%In this Letter, we investigate the polarization evolution of temporally incoherent light waves in dispersive and nonlinear optical fibers. This process is described by a {\it vector} system of two coherently coupled NLSEs. In this context, we will reveal the presence of two fundamentally different regimes: an irreversible thermalization, and a reversible turbulent dynamics driven by phase-correlations. 
Unlike recent 2D spatial studies in multimode optical fibers~\cite{podivilov22,PRL20,pourbeyram22,mangini22}, %where the propagation of coherent pulses was considered, 
%with quasi-monochromatic light \cite{ferraro23,ferraro25_rev}, 
we examine here the 
%longitudinal 
1D temporal dynamics of incoherent waves propagating in a single-mode fiber~\cite{babin07,turitsyn13,churkin15,PRL10,PR14}. 
%The main difference with respect to the recent studies in multimode optical fibers with (quasi-)monochromatic light \cite{PRA11,PRL19,pod19,PRA19,christodoulides19,kottos20,fan22,pod22,PRL22,PRL20,EPL21,wise_NP22,mangini22,PRL23,wright22}, is that we consider  the longitudinal temporal dynamics of an incoherent field that propagates in a single mode fiber \cite{PRL06,OE09,PRL10,OE11,turitsyn13,churkin15,PRX17}. 
By using the standard WT theory (neglecting phase-correlations), we 
%derive an effective kinetic equation that 
describe nonequilibrium thermalization accurately.
%which is characterized by the spontaneous repolarization of the temporally incoherent optical wave.
In contrast to this expected irreversible process, we show that phase-correlations can grow exponentially from fully uncorrelated initial random waves, leading to a reversible oscillatory turbulent dynamics.
These anomalous phase-correlations emerge among different wave components in a phase-invariant vector system, hence they are of different nature from those investigated in Ref.\cite{villois25}.
%which introduce a fundamental different behavior characterized by a reversible oscillatory turbulent dynamics. 
Experimental results provide evidence of both irreversible thermalization and reversible phase-correlated dynamics. From a broader perspective, this work advances our understanding of far-from-equilibrium self-organization processes in closed (Hamiltonian) turbulent wave systems.
%, with natural implications in nonlinear optics, hydrodynamics, plasma physics, or Bose-Einstein condensates.

{\it Model.}
We consider a generic model for vector phenomena based on coherently coupled NLSEs. It describes spinor Bose–Einstein condensates with 
field-induced 
spin-flip coupling \cite{stringari16}, allowing studies of 
%Faraday patterns \cite{cominotti22}, 
%paramagnetic to ferromagnetic 
quantum phase transitions \cite{cominotti23}, or bubble nucleation in 
ferromagnetic superfluids \cite{zenesini23}.
%It describes spinor Bose–Einstein condensates (BECs) with coherent coupling induced by an external field that drives spin-flip transitions. Such systems are key for exploring phenomena like Faraday pattern excitations \cite{cominotti22}, quantum phase transitions from paramagnetic to ferromagnetic states \cite{cominotti23}, and metastable bubble nucleation in ferromagnetic superfluids \cite{zenesini23}. 
%These studies promote coherently coupled BECs as an ideal platform to investigate out-of-equilibrium quantum field phenomena.
The coherently coupled NLSEs also describe the polarization dynamics of an optical wave propagating in  a weakly birefringent fiber \cite{millot_prl97,agrawal}: 
%In the circular polarization basis, the NLS model reads
%\begin{eqnarray}
%i\partial_z u_+ &=& - \beta \partial_{tt} u_+ + \alpha u_- + \gamma (|u_+|^2+ \kappa |u_-|^2)u_+ ,
%\label{eq:nls_p}\\
%i\partial_z u_- &=& - \beta \partial_{tt} u_- + \alpha u_+ + \gamma(|u_-|^2+ \kappa |u_+|^2)u_- ,
%\label{eq:nls_m}
%\end{eqnarray}
%\begin{eqnarray}
%i\partial_z u_x &=& - \beta \partial_{tt} u_x + \alpha u_x + 
%\gamma (|u_x|^2+ \kappa |u_y|^2)u + \rho u_x^* u_y^2,
%\label{eq:nlsu}\\
%i\partial_z u_y  &=& - \beta \partial_{tt} u_y - \alpha u_y + \gamma (|u_y|^2+ \kappa |u_x|^2)v  +  \rho u_y^* u_x^2,
%\label{eq:nlsv}
%\end{eqnarray}
\begin{eqnarray}
i\partial_z {\bm u} = - \beta \partial_{tt} {\bm u} + \alpha {\bm \sigma} {\bm u} +\gamma\big(\kappa {\bm u}^\dag {\bm u} {\bm u}+ \rho {\bm u}^T {\bm u} {\bm u}^*\big),
\label{eq:nls}
\end{eqnarray}
%with ${\cal F}({\bm u})=\kappa {\bm u}^\dag {\bm u} {\bm u}+(1-\kappa){\bm u}^T {\bm u} {\bm u}^*$, 
%plasma \cite{vnls_plasma}, hydrodynamics \cite{vnls_hydro} or Bose-Einstein condensates \cite{vnls_bec}:
%\begin{eqnarray}
%i\partial_z u &=& -  \partial_{tt} u + \alpha u + (|u|^2+ \kappa |v|^2)u + \rho u^* v^2,
%\label{eq:nlsu}\\
%i\partial_z v  &=& - \partial_{tt} v - \alpha v + (|v|^2+ \kappa |u|^2)v  + \rho v^* u^2,
%\label{eq:nlsv}
%\end{eqnarray}
where ${\bm u}(t,z)=(u_{x}, u_{y})^T$ is the vector field in the linear polarization basis, and the superscripts $(T,\dag)$ stand for the transpose, and conjugate transpose, operations. As usual in optics, the distance $z$ of propagation plays the role of an evolution `time' variable, while $t$ denotes the retarded time in a reference frame moving with the waves. 
%For convenience, we have written the vector NLS equation in dimensionless form, i.e., we have normalized the problem with respect to the nonlinear length $L_{nl}=1/(\gamma N/2)$, and the `healing' time, $\tau_0=\sqrt{\beta_2 L_{nl}/2}$, where $\gamma$ is the nonlinear coefficient, $\beta_2$ the dispersion coefficient, and $N$ is the average power of the waves \cite{supplemental}.
$\gamma$ is the non-linear parameter, $\beta$ is the group-velocity dispersion, while $\kappa$ and $\rho$ are dimensionless interaction coefficients; we will consider the case $\kappa=2\rho$ relevant to our experiments.
% and $\rho=\gamma/3$ the resonant nonlinear coupling.
%while the last term in  Eqs.(\ref{eq:nlsu}-\ref{eq:nlsv}) describes the resonant interaction between the two waves with the coefficient $\rho$. 
The coherent coupling parameter $\alpha$ originates from the weak birefringence of the optical fiber, with the matrix ${\bm \sigma}={\rm diag}(+1,-1)$. We assume $\alpha>0$ without loss of generality.
%The parameter $\alpha=\Delta \beta/2$ is the coherent coupling, with the matrix ${\bm \sigma}={\rm diag}(+1,-1)$, where $\Delta \beta=\beta_{0x}-\beta_{0y}$ accounts for a weak fiber birefringence, with $\beta_{0\mu}=n_\mu \omega_0/c$ ($\mu=x,y$), $\omega_0$ being the carrier optical frequency and $c$ the speed of light \cite{agrawal}. We assume $\alpha>0$ without loss of generality.
%with $T$ much larger than any characteristic length scale of the problem.
%The vector NLS Eq.(\ref{eq:nlsu}-\ref{eq:nlsv}) is known to model light propagation in weakly birefringent optical fibers, for which $\kappa=2/3$ and $\rho=1/3$.
%In this case, the component $u$ ($v$) denotes the $x$ ($y$) compo
%The dispersion relations of the waves then read
%\begin{eqnarray}
%k_{u}(\omega) = \omega^2 + \alpha, \quad k_{v}(\omega) = \omega^2 - \alpha.
%\end{eqnarray}
%In the numerical simulations we consider periodic boundary conditions over the numerical temporal window $[0,T_0]$. 
The vector NLSE (\ref{eq:nls}) conserves the power (particle number) $N=\sum_\mu N_{\mu}$, where $N_{\mu}(z)=\frac{1}{T}\int_0^T |u_{\mu}|^2 dt$, with $T$ the size of the numerical window, and $\mu=x,y$. It also conserves the Hamiltonian $H = E+U$, with a linear $E$, and a nonlinear $U$, contribution~\cite{supplement}.

Let us introduce the usual normal correlators $n_{\mu}(\omega,t,z)=\int \left<u_\mu(t+\tau/2,z)u_\mu^*(t-\tau/2,z)\right> \exp(-i\omega \tau)d\tau$ ($\mu=x, y$). The dependence of $n_\mu$ on the (`spatial') variable $t$ accounts for possible statistical inhomogeneities in the random waves, while the angle brackets denote an average over the random initial conditions. Furthermore, the theorem in Ref.\cite{villois25} for phase-invariant scalar systems can be extended to vector systems such as the vector NLSE (\ref{eq:nls}), which implies the absence of anomalous correlations among distinct frequencies. However, as we show below, phase-correlations may emerge between different  wave (i.e., polarization) components at the same frequency: these will be characterized by the anomalous correlator $m(\omega,t,z)=\int \left<u_y(t+\tau/2,z)u_x^*(t-\tau/2,z)\right> \exp(-i\omega \tau)d\tau$.
%according to Ref.\cite{villois25}, the system (\ref{eq:nls}) is phase-invariant, implying the absence of anomalous correlations among distinct frequencies. 

\medskip
\noindent
{\it Standard WT theory: Irreversible thermalization.} 
We consider the weakly nonlinear regime, where linear effects dominate over nonlinear effects  $|E/U| \gg 1$. Using the standard WT theory \cite{zakharov92,zakharov04,nazarenko11,Newell_Rumpf,galtier22,laurie12,onorato23,PR14}, we derive the KE under the usual assumptions: (i) statistical homogeneity, so that $n_\mu$ are $t-$independent, and (ii) absence of phase-correlations between $u_x$ and $u_y$, i.e., the anomalous correlator is zero at any `time' $z$, $m(\omega,t,z)=0$. 
%Assuming furthermore that the statistics is homogeneous, the normal correlators $n_{\mu}(\omega,z)$
%$n_{\mu}(\omega,t,z)=\int \left<u_\mu(t+\tau/2,z)u_\mu^*(t-\tau/2,z)\right> \exp(-i\omega \tau)d\tau$ ($\mu=x, y$)
%do not depend on the (`space') variable $t$,  
%$n_{\mu}(\omega,t,z) \to n_{\mu}(\omega,z)$, 
The KE governing the evolution of the averaged spectra ${\bm n}(\omega,z)=(n_x, n_y)^T$ then takes the usual form
\begin{align}
\partial_z {\bm n}(\omega,z) = \kappa^2 \, {\cal C}oll_i[{\bm n}] + \rho^2 \, {\cal C}oll_c[{\bm n}].
\label{eq:wtkin} 
\end{align}
The collision terms are rather cumbersome, see \cite{supplement}. They are cubic nonlinear terms ${\cal C}oll_{i,c}[{\bm n}] \sim n^3$, describing the four-wave interaction as a collisional gas of `particles'. The KE (\ref{eq:wtkin}) conserves 
%the total number of particles, 
$N=\sum_\mu N_\mu$, with ${N}_\mu(z)= \frac{1}{2\pi}\int {n}_\mu(\omega,z) d\omega$ ($\mu=x,y$), and the linear energy $E$ \cite{supplement}.
%${E}= \sum_\vfi {E}_\mu$, with ${E}_\vfi=\frac{1}{2\pi}\int k_\vfi(\omega) {n}_\vfi(\omega,z) d\omega$, with $k_{u_{x,y}}(\omega)=\omega^2 \pm \alpha$. 
%Here, we only describe their main properties. First of all, they are cubic nonlinear terms, ${\cal C}oll_{i,c}[{\bm n}] \sim n^3$. The collision term ${\cal C}oll_i[{\bm n}]$ describes the incoherent interaction as a collisional gas of `particles' polarized along $x$ and $y$, $(x,y) \to (x,y)$. 
%Accordingly, ${\cal C}oll_i{\bm n}]$ conserves the number of `particles' $N_{x}$ and $N_{y}$ in each polarization component. ${\cal C}oll_i[{\bm n}]$ describing the coherent coupling refers to a process in which two particles polarized along $x$ are destroyed, while two particles polarized along $y$ are created, $(x,x) \to (y,y)$, and vice-versa. Accordingly, the KE (\ref{eq:wtkin}) 
%is responsible for an exchange of particles among $u_x$ and $u_y$, so that 
%$N_{u_x,u_y} \neq$const, which in turn leads to the effect of repolarization.
%the KE (\ref{eq:wtkin_u}) only conserves the total number of particles, $N=\sum_\mu N_\mu$ with ${N}_\mu= \frac{1}{2\pi}\int {n}_\mu(\omega,z) d\omega$ ($\mu=u_x,u_y$). 
%the total momentum, ${\tilde P}=\sum_\vfi {\tilde P}_\vfi$, ${\tilde P}_\vfi=\int k {\tilde n}_\vfi(k,t) dk$ 
At variance with the NLSE~(\ref{eq:nls}), the KE (\ref{eq:wtkin}) is irreversible, as expressed by a $H$-theorem of entropy growth, $d{\cal S}/dz \geq 0$, where ${\cal S}=\sum_\mu {\cal S}_\mu$, and ${\cal S}_{\mu}(z)=\frac{1}{2\pi}\int \log\big({n}_\mu(\omega,z)\big) d\omega$  is the nonequilibrium entropy of the $\mu-$th component ($\mu=x,y$). 
%Also note that the RJ thermodynamic spectra ${n}_{u_x,u_y}^{RJ}(\omega)$ are obtained by maximizing the entropy ${{\cal S}}[{n}_u,{n}_v]$ under the constraints of conservation of ${E}$ and $N$, see \cite{supplement}. However, the RJ spectra are physically irrelevant because they lead to a divergence of the linear energy, $E$.
%${n}_{u_x}^{RJ}(\omega)=\frac{T}{\omega^2 +\alpha - {\nu}}$, 
%${n}_{u_y}^{RJ}(\omega)=\frac{T}{\omega^2 -\alpha - {\nu}}$, where $T$ (`temperature'), and ${\nu}$ (`chemical potentials') are the Lagrangian multipliers associated to the conservation of ${E}, {N}$, respectively.
%Note in particular that the RJ distribution (\ref{eq:neqRJ}) is a stationary solution of the kinetic Eqs.(\ref{eq:wtkin_u}).

\begin{figure}[t!]
    \centering
    \includegraphics[trim = 0.5cm 0 0 0,width=1\linewidth]{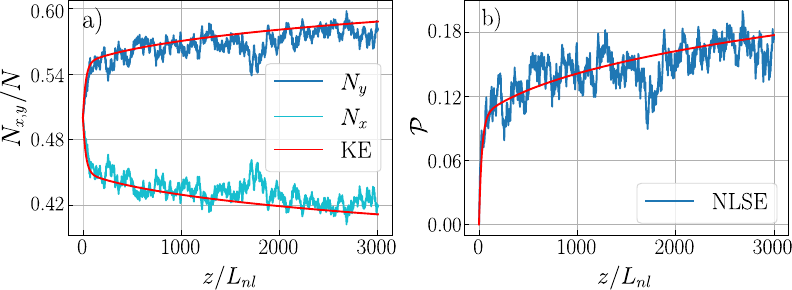}
    \caption{{\bf Irreversible thermalization.} Evolution during the propagation of \blue{of the power fraction $N_{x,y}(z)/N$} (a), and corresponding degree of polarization ${\cal P}(z)$ [Eq.(\ref{eq:P})] (b), obtained from the simulation of the NLSE (\ref{eq:nls}) (5 realizations, dark and light blue lines), and the simulation of the WT KE (\ref{eq:wtkin}) (red lines). 
    %The thermalization process is characterized by the repolarization of the random wave. 
%    Phase-correlations do not grow, $M(z) \simeq 0$, and ${\cal P}(z) \simeq \Delta N(z)/N$.
    Parameters: $\alpha L_{nl}=1, \Delta N^0=0, \sigma \tau_0=0.8 \pi, \kappa=2/3$, with $L_{nl}=1/(\gamma N)$ the nonlinear length and $\tau_0=\sqrt{|\beta| L_{nl}}$ the `healing time' ($|E/U| \simeq (\sigma \tau_0)^2$).
    %$\sigma=0.7$
    }
    \label{fig:fig1}
\end{figure}

%In the Supplementary Material, we derive a reduced form of the KE (\ref{eq:wtkin}) (see Eqs.(??) in \cite{supplement}), which we integrate numerically to compare its results with those of the NLSE (\ref{eq:nls}).
%Without loss of generality, we assume hereafter $\alpha >0$.
We have performed numerical simulations of NLSE (\ref{eq:nls}).
%and WT KE (\ref{eq:wtkin}). 
As initial condition, we consider two {\it uncorrelated} random waves $u_{x,y}(t,z=0)$ with zero mean, Gaussian statistics and Gaussian-shaped initial spectrum $n_{x}^0(\omega)=n_{y}^0(\omega)$, i.e., $N_x^0=N_y^0$, with $n_{\mu}^0(\omega)= \sqrt{2\pi} N_{\mu}^0 \exp\big(-\omega^2/(2\sigma^2)\big)/\sigma$.
Let us introduce the degree of polarization of the optical wave \cite{mandel}:
\begin{eqnarray}
{\cal P}(z)=\sqrt{\Delta N(z)^2 + 4|M(z)|^2 }/N,
\label{eq:P}
\end{eqnarray}
where $M(z)=\frac{1}{2\pi}\int m(\omega,z) d\omega = \left< u_y(t,z) u_x^*(t,z) \right>$ denotes the integrated anomalous correlator, and $\Delta N(z)=N_y(z)-N_x(z)$ the unbalanced power distribution. Note that ${\cal P}$ is bounded by 0 and 1, which correspond to unpolarized, and fully polarized, waves, respectively. 

We have also performed numerical simulations of the WT KE~(\ref{eq:wtkin}): as shown in Fig.~\ref{fig:fig1}, a good agreement (without adjustable parameters) is obtained with direct simulations of the NLSE (\ref{eq:nls}). 
%Recalling that the KE neglects phase-correlations, such an agreement means that anomalous correlations do not grow, i.e., $M(z) \simeq 0$, and ${\cal P}(z) \simeq \Delta N(z)/N$. Accordingly, the thermalization process is characterized by an irreversible transfer of power from $N_x$ to $N_y$ due to their opposite contributions to the linear energy, thus leading to the emergence of a non-vanishing degree of polarization \cite{supplement}. 
Recalling that the KE neglects phase-correlations, such an agreement means that anomalous correlations do not grow, i.e., $M(z) \simeq 0$ and ${\cal P}(z) \simeq \Delta N(z)/N$. Accordingly, the thermalization process is characterized by an irreversible transfer of power from $N_x$ to $N_y$, 
%due to their opposite contributions to the linear energy, 
thus leading to the emergence of a non-vanishing degree of polarization \cite{supplement}. 
\medskip
\noindent
{\it Reversible dynamics mediated by phase-correlations.} 
%Simulations of the NLSE (\ref{eq:nls})  unexpectedly reveal that, in certain cases, the thermalization process is arrested by the spontaneous emergence of phase  correlations. 
We now show that the thermalization process can be inhibited by the spontaneous emergence of phase-correlations. To clarify this regime in a general framework, we do not implicitly assume the waves to obey homogeneous statistics, i.e., we leave $n_{\mu}(\omega,t,z)$ and $m(\omega,t,z)$ to depend on a slow non-homogeneous variation in the $t$ variable.
%`space' variable $t$. 
Starting from the NLSE (\ref{eq:nls}), we derive in the Supplementary Mayerial  the equations governing the coupled evolution of the normal and anomalous correlators \cite{supplement}:
\begin{widetext}
\begin{align}
\partial_z n_x(\omega,t) =& - 2\beta \omega \partial_t n_x 
+ \gamma \partial_t \big(2 N_x(t) + \kappa N_y(t) \big)\partial_\omega n_x
+2\gamma\kappa \partial_t M^r(t) \partial_\omega m^r
+4 \gamma\kappa M^r(t) m^i  ,
\label{eq:nx_vl} \\
%\partial_z n_y(\omega,t) =& - 2 \beta\omega \partial_t n_y + \gamma \partial_t \big(2 N_y(t) + \kappa N_x(t) \big)\partial_\omega n_y
%+2\gamma\kappa \partial_t M^r(t) \partial_\omega m^r(\omega,t)
%-4 \gamma\kappa M^r(t) m^i(\omega,t)
%\label{eq:ny_vl} \\
\partial_z m(\omega,t) =& -2\beta \omega \partial_t m 
+\gamma(1+\kappa/2) \partial_t \big( N_x(t)+N_y(t) \big) \partial_\omega m
+\gamma\kappa \partial_t M^r(t)\partial_\omega \big( n_x+n_y \big)
\nonumber \\
&-i m \Big( \gamma(2-\kappa)\big(N_y(t)-N_x(t)\big)-2\alpha   \Big)  
- 2i\gamma\kappa \big( n_x - n_y \big) M^r(t)   ,
\label{eq:m_vl}
\end{align}
\end{widetext}
with ${M}(t,z)=\frac{1}{2\pi}\int m(\omega,t,z)d\omega$, ${N}_{\mu}(t,z)=\frac{1}{2\pi}\int n_{\mu}(\omega,t,z)d\omega$, while the superscripts $m^{r,i}$ ($M^{r,i}$) denote the real and imaginary parts of $m$ ($M$). 
The equation for $n_y$ follows from Eq.(\ref{eq:nx_vl}) by exchanging $x\leftrightarrow y$ and reversing the sign of the last term.

The first two terms on the right-hand side of Eq.(\ref{eq:nx_vl}) 
%Eqs.(\ref{eq:nx_vl}-\ref{eq:ny_vl}) 
correspond to the well-established Vlasov coupling between two {\it uncorrelated} random waves. Here, the generalized Vlasov-like Eqs.(\ref{eq:nx_vl}-\ref{eq:m_vl}) provide an original extension accounting for phase-correlations, $m(\omega,t)$. In contrast to the KE (\ref{eq:wtkin}), the system (\ref{eq:nx_vl}-\ref{eq:m_vl}) is formally reversible in `time' $z$ \cite{supplement}. To avoid confusion with the usual KE~(\ref{eq:wtkin}), we shall refer to Eqs.(\ref{eq:nx_vl}-\ref{eq:m_vl}) (and Eq.(\ref{eq:s_dyn}) below) as the anomalous-correlator kinetic equation (AC-KE).

\begin{figure}
    \centering
    \includegraphics[trim = 0.3cm 0 0 0,width=\linewidth]{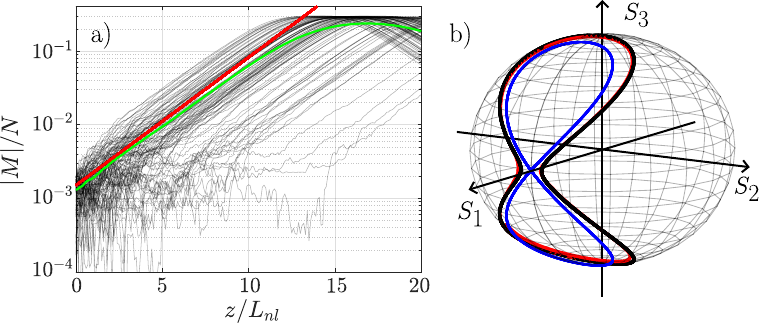}
    \caption{{\bf Reversible turbulent dynamics.} (a) Exponential growth of the anomalous correlator starting from initially uncorrelated random waves $|M(z=0)| \simeq 0$: Simulations of the NLSE (\ref{eq:nls}) (100 realizations, black lines), and corresponding average (green line). The red line reports the theoretical prediction Eq.(\ref{eq:lambda_0}). (b) Nonlinear dynamics on the Poincar\'e sphere: Simulation of the NLSE (\ref{eq:nls}) (red  line), and corresponding theoretical prediction from the simulation of AC-KE~(\ref{eq:s_dyn}) (black line), showing the periodic exchange among the normal and anomalous correlators. The blue line reports the homoclinic orbit emanating from the unstable fixed point $S_1=+S_0$, i.e., $|M|=0$. Parameters: $\alpha L_{nl}=0.2, \sigma \tau_0=4\pi, \Delta N^0/N=0.6, \kappa=2/3, L_{nl}=1/(\gamma N), \tau_0=\sqrt{|\beta| L_{nl}}$ ($|E/U| \simeq (\sigma \tau_0)^2$).
    %$\sigma=4$
    }
    \label{fig:fig2}
\end{figure}

We consider an initial condition of two uncorrelated random waves, of zero mean, with homogeneous statistics and spectra $n_{\mu}(\omega,z=0)=n_{\mu}^0(\omega)$, so that the initial anomalous correlator is zero, $m(\omega,t,z=0)=0$. We carry out a linear stability analysis of Eqs.(\ref{eq:nx_vl}-\ref{eq:m_vl}) around this state. Using the Laplace-Fourier transform ${\hat m}(\omega, \Omega,\lambda)=\int_0^z dz \int m(\omega,t,z)\exp(-\lambda z-i\Omega t) dt$, we obtain the dispersion relation %for the growth rate 
$\lambda(\Omega)$ of the anomalous correlator \cite{supplement}: 
\begin{eqnarray*}
%\frac{2\pi}{\gamma\kappa} = \sum_{s} \int  
%\frac{ \big(n_x^0-n_y^0\big)(\omega)-s\Omega \partial_\omega\big(n_x^{0}+n_y^{0}\big)(\omega)/2}{is \lambda-2s\beta\omega \Omega - \gamma(2-\kappa)\Delta N^0+2\alpha} d\omega  ,
\frac{2\pi}{\gamma\kappa} = \sum_{s} \int  
\frac{ n_x^0(\omega)-n_y^0(\omega)-s\Omega \partial_\omega\big(n_x^{0}(\omega)+n_y^{0}(\omega)\big)/2}{is \lambda-2s\beta\omega \Omega - \gamma(2-\kappa)\Delta N^0+2\alpha} d\omega  ,
%\label{eq:disp_rel}
\end{eqnarray*}
where $\Delta N^0=N_y^0-N_x^0$, and $s=\pm 1$.
%and  the sum is carried over $s=\pm 1$.
Assuming the initial spectra Gaussian-shaped, 
%(as in Fig.~\ref{fig:fig1}), 
%$n_{\mu}^0(\omega)= \sqrt{2\pi} N_{\mu}^0 \exp\big(-\omega^2/(2\sigma^2)\big)/\sigma$, 
we compute the corresponding growth-rate ${\rm Re}[\lambda(\Omega)]$.
%solution of Eq.(\ref{eq:disp_rel}). 
The analysis reveals that, in general, the homogeneous mode with $\Omega=0$ is the most unstable, with the maximum growth rate of the anomalous correlator: 
%$\lambda_0=\lambda(\Omega=0)$:
%for anomalous correlations is 
\begin{align}
\lambda(\Omega=0)=  2\sqrt{\alpha(2 \gamma \Delta N^0/3-\alpha )},
%\lambda(\Omega=0)=  
% \sqrt{\big(\gamma(2-\kappa)\Delta N^0-2\alpha\big)\big( \gamma(3\kappa-2)\Delta N^0+2\alpha\big)}.
\label{eq:lambda_0}
\end{align}
where we have considered the case $\kappa=2/3$ relevant to our experiments.
%Considering the case $\kappa=2/3$ relevant to our experiments, 
%phase-correlations between the initial uncorrelated random waves $u_{x,y}$ then grow exponentially when 
The instability criterion then reads 
\begin{align}
\alpha L_{nl} < (2/3) \Delta N^0/N.
\label{eq:threshold}
\end{align}
%$\alpha L_{nl} < (2/3) \Delta N^0/N$,
%to save place: revealing that the emergence of phase-correlations requires an unbalanced initial population, $N_y^0 > N_x^0$. 
%The simulations of the NLSE (\ref{eq:nls}) confirm the theoretical prediction (\ref{eq:lambda_0}), 
Simulations of NLSE [Eq. (\ref{eq:nls})] confirm the theoretical prediction [Eq. (\ref{eq:lambda_0})], \blue{as illustrated in Fig. \ref{fig:fig2}(a). The figure shows 100 independent realizations of the initial uncorrelated random waves $u_{x,y}(t,z=0)$ (black lines); whose ensemble average (green line) agrees with the instability growth-rate (\ref{eq:lambda_0}) (red line).}
%-- note that the above instability condition is not verified in Fig.~1 (since $\alpha L_{nl}=1$).

The fact that the homogeneous mode $\Omega=0$ is the most unstable, indicates that the subsequent nonlinear dynamics described by Eqs.(\ref{eq:nx_vl}-\ref{eq:m_vl}) preserves the homogeneous statistics, i.e., $n_{\mu}(\omega,t)$ and $m(\omega,t)$ are $t-$independent. In this limit, the  dynamics for the normal and anomalous correlators (\ref{eq:nx_vl}-\ref{eq:m_vl}) can be recast in the form of a Stokes vector ${\bm S}(z)=\big(\Delta N, 2 M^r, 2 M^i \big)^T$ that rotates on the surface of the Poincar\'e sphere \cite{supplement}: 
\begin{align}
\partial_z {\bm S}(\tau,z)= {\bm R}(z) \times  {\bm S}(z),
\label{eq:s_dyn}
%&{\bm R}(z)=\big( 2\alpha-\blue{\gamma}(2-\kappa) S_1(z), \ -2\blue{\gamma}\kappa S_2(z), \ 0  \big).\label{eq:R}
\end{align}
with ${\bm R}(z)=\big( 2\alpha-\blue{\gamma}(2-\kappa) S_1(z),  -2\blue{\gamma}\kappa S_2(z), 0  \big)^T$.
This formalism differs substantially from that commonly used to describe polarization effects for fully coherent stationary waves \cite{agrawal,daino85}.
\blue{Conversely, a related kinetic approach for random waves accounting for anomalous correlations was originally developed in the context of isotropic plasma turbulence, where nonlinear interactions due to stimulated scattering drive the electromagnetic field toward complete polarization \cite{kuznetsov78} (we recall that classical Thomson scattering %of unpolarized light 
likewise produces fully polarized light at a scattering angle $\pi/2$ \cite{landau}). 
Note however that the increase of ${\cal P}(z)$ discussed above through Fig.~\ref{fig:fig1} is of different nature, since it is associated with the thermalization process that neglects anomalous correlations. Furthermore, at variance with Ref.\cite{kuznetsov78}, the kinetic Eq.(\ref{eq:s_dyn}) accounting for anomalous correlators, conserves the degree of polarization during evolution ${\cal P}(z)=$const, as it is fixed by the radius of the Poincar\'e sphere $S_0=\big(\sum_{j=1}^3 S_j^2\big)^{1/2}={\cal P} N$. It turns out that} the nonlinear dynamics described by the AC-KE~(\ref{eq:s_dyn}) is essentially periodic, featuring oscillations that involve a reversible exchange between the normal correlator $\Delta N(z)$ and the anomalous correlator $M(z)$, while conserving
\begin{align}
{\cal P}^2 N^2 =\Delta N^2(z)+4|M(z)|^2={\rm const}.
\label{eq:invar}
\end{align}
This reversible dynamics is in agreement with the simulations of the NLSE (\ref{eq:nls}), see Fig.~\ref{fig:fig2}(b). 
Note that the {\it fast}, reversible dynamics, governed by the quadratic nonlinearities 
%involving the anomalous correlator 
in the AC-KE~(\ref{eq:nx_vl}-\ref{eq:m_vl}) and (\ref{eq:s_dyn}) in Fig.~\ref{fig:fig2}, stands in contrast to the {\it slow}, irreversible thermalization driven by the cubic nonlinear KE~(\ref{eq:wtkin}) in Fig.~\ref{fig:fig1}.
%It should be noted that the characteristic timescale of the dynamics in Fig.2 is on the order of $\sim 10L_{nl}$, in sharp contrast to the significantly slower thermalization timescale presented in Fig.~1. The rapid dynamics associated with the anomalous correlator originate from the quadratic nonlinearity of Eqs. (..), whereas the slow process of dynamical thermalization is governed by the cubic nonlinearity of the kinetic equations...

\medskip
\noindent
{\it Experiments.}
We have experimentally investigated the two different turbulent regimes that we have previously described. We use $100$-ps long pulses made of temporally incoherent optical waves with a near Gaussian beam shape and the following spectro-temporal features: $558$-nm central wavelength, full width at half maximum (FWHM) of about $1.93$ THz~\cite{supplement}.
The temporally incoherent pulse is then divided into two linear polarization states with relative tunable power and temporal delay (much larger than their coherence time). The incoherent waves are then injected into a 6.2-m-long weakly birefringent 
%spun 
silica fiber ($\alpha \approx 0.565 \ \rm{m}^{-1}$ at $558\ \rm{nm}$) 
%($\Delta \beta \approx 1.13 \ \rm{m}^{-1}$ at $558\ \rm{nm}$), 
whose propagation is governed by Eq.(\ref{eq:nls})~\cite{agrawal}. The complex anomalous correlator $M$, as well as the optical powers $N_{x,y}$ and spectra, along $x$ and $y$ axes, are measured at the input and output of the fiber using an optical spectrum analyzer and a polarimeter.

\begin{figure}[t!]
    \centering
    \includegraphics[trim = 0.5cm 0cm 0cm 0cm, width = \linewidth]{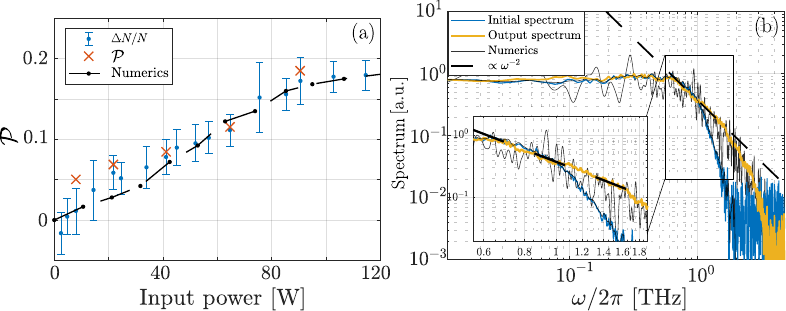}
    \caption{\textbf{Observation of optical repolarization.} (a) Experimental measurements of the normalized power difference $\Delta N/N$ (blue dots), and degree of polarization ${\cal P}$  (red crosses), vs input power: In the thermalization regime, the anomalous correlator $M$ is negligible, and ${\cal P} \simeq \Delta N/N$, see Eq.(\ref{eq:P}). The measured repolarization is in good agreement with the simulations of NLSE~(\ref{eq:nls}) (dashed black lines). (b) Measured power spectrum at the input (blue line), and output (orange line), of the fiber. The experimental spectra agree well with 
    %those obtained from 
    the simulation of NLSE~(\ref{eq:nls}) (black lines), and exhibit a power-law signature of light thermalization $\sim \omega^{-2}$ in the spectral tail.}
    \label{fig:fig3}
\end{figure}

\noindent
\blue{\it{Experiments on the thermalization regime.}} 
%We begin by investigating the irreversible thermalization process. 
This regime is obtained when injecting the same power along both axes $x$ and $y$, i.e., $\Delta N^0 =0$.
%, for which the anomalous correlator is stable. 
Fig.~\ref{fig:fig3}(a) shows the measurements of the power difference $\Delta N/N$ and degree of polarization ${\cal P}$ at the fiber output, as a function of the input power. We observe that the degree of polarization is effectively determined by the power imbalance ${\cal P} \simeq \Delta N/N$ (see Eq.(\ref{eq:P})), since the measured anomalous correlator is negligible ($|M| \simeq 0$), as expected for the thermalization regime under investigation.
%We have checked this by directly measuring the phase-correlations and found $M\approx 0$, so that ${\cal P} \simeq \Delta N/N$ from Eq.~(\ref{eq:P}). 
%Fig.\ref{fig:fig1}-a) shows the power difference on the fast and slow axes as a function of the input power. 
%\blue{\it Tu peux verifier que $\alpha L_{nl} > (2/3) \Delta N^0/N$ meme pour de grandes puissances?}\red{Tu veux dire en fin de propagation ? Parce que $\Delta N^0 = 0$ par construction non?} \blue{oui, je voulais dire avec $\Delta N^0$ en sortie.}\red{J'ai vérifié et on a toujours $\alpha L_{nl} > (2/3) \Delta N^0/N$, pour la plus grande puissance en sortie de fibre on a $\alpha L_{nl} \approx 0.19$ et $(2/3)\Delta N^0/N\approx 0.12$.} 
In Fig.~\ref{fig:fig3}(a) we can see that, as the input power increases, the degree of polarization ${\cal P}$ at the fiber output 
%also 
grows larger. This behavior is found to be in agreement with the numerical simulations of NLSE~(\ref{eq:nls}), performed by faithfully reproducing the details of the experimental conditions \cite{supplement}. As predicted by \blue{the criterion (\ref{eq:threshold})}, a sufficiently large power difference may render the anomalous correlator unstable. It is worth emphasizing, however, that even for the largest power difference recorded at the fiber output in Fig.~\ref{fig:fig3}(a), the anomalous correlator remains stable (which is consistent with \blue{the criterion (\ref{eq:threshold})}, as $\alpha L_{nl} \approx 0.19 > (2/3)\Delta N/N \approx 0.12$).

Figure~\ref{fig:fig3}(b) shows the measured input and output spectra along the $x-$axis of the fiber (similar results are obtained along the $y-$axis), together with the spectrum obtained from NLSE simulations. The results show the formation of a spectral tail decaying with the power law $\sim \omega^{-2}$, which is a signature of light thermalization reflecting energy equipartition among the modes~\cite{zakharov92,zakharov04,nazarenko11,galtier22,Newell_Rumpf,PR14}.
%(also see Fig.~S4 in \cite{supplement}).
%Note that a significant part of the spectral tail decays with the power law $\sim \omega^{-2}$,  reflecting an energy equipartition among the modes~\cite{zakharov92,Newell01,nazarenko11,galtier22,Newell_Rumpf,PR14}. This signature of light thermalization was discussed in different works~\cite{OE09,PRL10,OE11,turitsyn13,churkin15}, but not observed so far.
\begin{figure*}[t!]
    \centering
    \includegraphics[trim = 0cm 0 0 0, width=\linewidth]{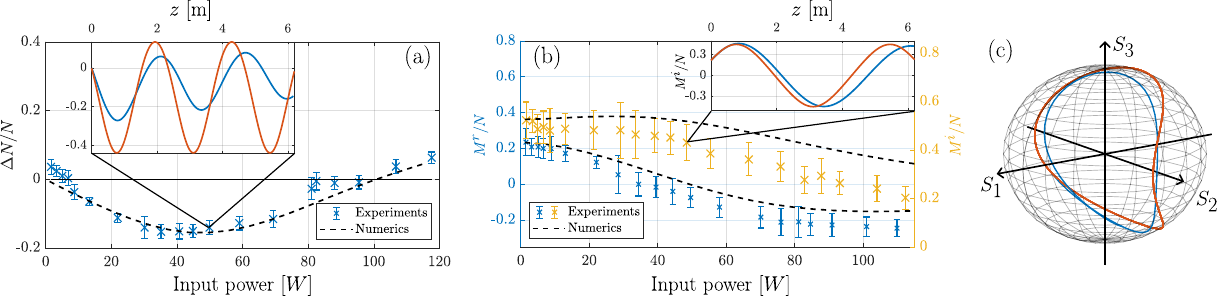}
\caption{\textbf{Observation of the reversible turbulent regime}. 
Measurements, at the fiber output and as a function of the input power, of (a) the power difference $\Delta N/N$ and (b) the real and imaginary parts of the normalized anomalous correlator $M/N$.
%(a) Evolution of the power difference $\Delta N/N$, at the fiber output as a function of the input power. 
%The inset shows the evolution along the fibre in the nonlinear regime, $E/U \approx 1$ for an input peak power of $50\ \rm{ W}$ ($L_{nl}\approx 0.83$ m). The blue curve depicts the NLSE evolution while the red curve shows the evolution according to the KE.  
%(b) Real and imaginary part of the normalized anomalous correlator $M/N$ as a function of the input power. 
%We see that both $\Delta N/N$ and $M/N$ qualitatively agree with the simulations of NLSE (\ref{eq:nls}) (dashed black line). 
%The inset shows the evolution along the fiber of the imaginary part of the anomalous correlator according to the NLSE (blue curve) and the KE (red curve) for $E/U \approx 1$. 
%Note that the both the amplitude and the period of oscillation are in qualitative agreement.
The insets in (a)-(b) show the dynamics along the fiber length $L=6.2$m for $\Delta N/N$ and $M^i/N$ with an input power of $50\ \rm{ W}$ ($L_{nl}\approx 0.83$ m) in the strongly nonlinear regime $|E/U| \approx 1$: The blue and red lines report the simulations of NLSE (\ref{eq:nls}) and AC-KE (\ref{eq:s_dyn}), respectively.    
(c) Corresponding field evolution on the Poincar\'e sphere 
%over the fiber length $L = 6.2\ \rm{m}$ 
for the NLSE (blue) and the AC-KE (\ref{eq:s_dyn}) (red).  
%according to the NLSE model (blue curve)
%, then extended to a length $L = 62\ \rm{m}\approx 75L_{nl}$ (dashed red curve) 
%and the periodic evolution predicted by the AC-KE (\ref{eq:s_dyn}) (red curve). 
%We can see that the KE conserves the degree of polarization ${\cal P}$, while the spiraling trajectory of the NLSE reflects a depolarisation effect.
Note in the insets (a)-(b) and in panel (c) that the AC-KE (\ref{eq:s_dyn}) (valid in the weakly nonlinear regime $|E/U| \gg 1$) still provides a qualitative description of the oscillatory dynamics even in the strongly nonlinear regime $|E/U| \approx 1$.
}
\label{fig:fig4}
\end{figure*}

\noindent
\blue{{\it Experiments on the reversible turbulent regime.} 
%We now consider the second regime characterized by a reversible oscillatory turbulent dynamics. 
For small initial anomalous correlations  $|M(z=0)| \simeq 0$, we have seen through the criterion (\ref{eq:threshold}) that $|M|$ can be unstable. Experimental constraints, however, prevent simultaneously launching a strong  unbalanced population $S_1 \simeq S_0$ with a small phase-correlation $|M|\simeq 0$, precluding a direct experimental test of criterion (\ref{eq:threshold}). 
Nevertheless, in addition to the instability of the fixed point $(M=0, S_1 = +S_0 )$, the AC-KE~(\ref{eq:s_dyn}) 
also predicts a general nonlinear coupling between normal and anomalous correlators for arbitrary initial conditions on the sphere ${\bm S}(z=0)$, even if the criterion (\ref{eq:threshold}) is not fulfilled. 
Here, we probe this nonlinear dynamics experimentally starting from $S_1(z=0) \simeq 0$ with a large phase correlation $|M|$.}
%We now turn our attention to the second regime characterized by a reversible oscillatory turbulent dynamics. Constraints on the injection conditions prevent simultaneously launching a strong  unbalanced population $S_1$ with a small phase-correlation $|M|\simeq 0$, so that we are unable to test the criterion (\ref{eq:threshold}). However, aside from the instability of the fixed point $S_1 = +S_0$, the AC-KE~(\ref{eq:s_dyn}) predicts a general nonlinear coupling between normal and anomalous correlators for arbitrary initial conditions on the sphere ${\bm S}(z=0)$. In our experiments, we probe this dynamics starting from $S_1(z=0) \simeq 0$ with a large phase-correlation $|M|$. 
We thus restore a large anomalous correlator by means of a polarizer, placed before injection of the laser beam into the fiber.
%At the output of the fiber, we measure  the power difference as well as the complex anomalous correlator. 
%Fig.~\ref{fig:fig4} shows the experimental results and corresponding simulations of the NLSE (\ref{eq:nls}). 
Figs.~\ref{fig:fig4}(a-b) report the experimental measurements of the power difference $\Delta N/N$ and complex anomalous correlator $M/N$ at the fiber output, as functions of the input power. Their non-monotonic evolutions agree well with the simulations of the NLSE~(\ref{eq:nls}). 
Such an agreement has been obtained by using a single adjustable parameter, namely the initial relative phase between $u_x$ and $u_y$, which accounts for a minor, uncontrolled elliptical polarization of the injected incoherent optical beam. 
\blue{We have deliberately chosen to limit the number of adjustable parameters in order to highlight the physical mechanism at play. The quantitative discrepancy between simulations and experiments in Fig.~\ref{fig:fig4}(b) can also be attributed to the phase-sensitive nature of the anomalous correlator $M/N$, which makes it more vulnerable to experimental uncertainties and noise than $\Delta N/N$ shown in Fig.~\ref{fig:fig4}(a).}
%The inset in Fig.~\ref{fig:fig4}(a-b) illustrate a periodic evolution along $z$ in the regime where linear and nonlinear effects are of the same order $E/U\approx1$ ($N=50\ \rm{W}$) and modeled by the NLSE. This is the remnant of the periodic evolution predicted in the weak nonlinear regime by Eq.~\ref{eq:s_dyn} (red curve in the inset). Similarly, the inset in Fig.~\ref{fig:fig4}(b) shows the periodic evolution of the imaginary part of the anomalous correlator modeled by the NLSE. Note that for the anomalous correlator both the phase and amplitude agree well with the KE. Fig.~\ref{fig:fig4}(c) shows the numerical evolution of the optical field on the Stokes sphere for $E/U\approx1$ The curves illustrate the periodic evolution in the weakly nonlinear regime describe by the Eq~(\ref{eq:s_dyn}) accounting for phase correlations and according to the NLSE.~(\ref{eq:nls}) (red line). 
%The analysis of the temporal $z$-evolution of the normal and anomalous correlators uncovers the underlying oscillatory turbulent dynamics. The insets of Fig.~\ref{fig:fig4}(a-b) show the evolution of $\Delta N/N$ and $M^i/N$ along the fiber length $L=6.2$m, for an input power of 50W, which corresponds to a strongly nonlinear regime with $|E/U| \simeq 1$. 
The  analysis of the evolution of $\Delta N/N$ and $M/N$ along the fiber length $L=6.2$m uncovers the underlying oscillatory turbulent dynamics. This is illustrated in the insets of Fig.~\ref{fig:fig4}(a-b) for an input power of 50W, which corresponds to a strongly nonlinear regime with $|E/U| \simeq 1$. 
The blue and red curves refer to simulations of the NLSE~(\ref{eq:nls}) and the AC-KE~(\ref{eq:s_dyn}), respectively, whose  corresponding evolution is also reported on the Stokes sphere, see Fig.~\ref{fig:fig4}(c).
These results show that the AC-KE~(\ref{eq:s_dyn}), which is rigorously valid in the weakly nonlinear regime $|E/U| \ll 1$ remains robust in describing qualitatively the oscillatory turbulent behavior even in the strongly nonlinear regime $|E/U| \simeq 1$.
%-- despite quantitative differences with NLSE predictions, the AC-KE~(\ref{eq:s_dyn}) still captures the essential qualitative oscillatory dynamics. 
%These results show that the AC-KE~(\ref{eq:s_dyn}), which is rigorously valid in the weakly nonlinear regime $|E/U| \ll 1$ remains robust in describing  the reversible turbulent behavior even in the strongly nonlinear regime $|E/U| \simeq 1$ -- despite quantitative differences with NLSE predictions, the AC-KE~(\ref{eq:s_dyn}) still captures the essential qualitative oscillatory dynamics. 
%Note in particular that, in contrast to the irreversible repolarization regime in Fig.~\ref{fig:fig3}, here the degree of polarization is conserved through propagation ${\cal P}(z) \simeq$const.
%, as the trajectory in Fig.~\ref{fig:fig4}(c) remains close to the surface of the Poincar\'e sphere.
%In the nonlinear regime and for long distances, $L\gg L_nl$, the degree of polarization ${\cal P}$ is not conserved during propagation -- the trajectory exhibits a spiraling pattern with a decreasing radius, reflecting the depolarization of the incoherent wave (shown in dashed red).

\medskip
\noindent
{\it Conclusion and perspectives.} 
We have revealed, both theoretically and experimentally, that a phase-invariant Hamiltonian system of coupled waves may exhibit two distinct  turbulent regimes:  irreversible thermalization and reversible dynamics mediated by phase-correlations. 
%The criterion (\ref{eq:threshold}) determines which regime emerges when the initial anomalous correlator vanishes, while a non-vanishing initial anomalous correlator drives the system into the reversible turbulent regime. 
The experiments provide 
%to the best of our knowledge, 
the first evidence of such contrasting turbulent behaviors in conservative (Hamiltonian) systems of dispersive nonlinear waves.
%, we have derived kinetic (Vlasov-like) equations describing the coupled evolution of normal and anomalous correlators. 
%We have  reported an experimental system that enables the investigation of conservative nonlinear regimes exhibiting both irreversible thermalization and reversible dynamics mediated by phase-correlations. 

Our study establishes a basis for future advances in understanding the interplay of these fundamentally different regimes. Until now, turbulent regimes dominated by slow, irreversible thermalization and those governed by fast, correlation-driven reversible dynamics have been separately explored, leaving open the fundamental question about the role of anomalous correlators on thermalization, and the reciprocal impact of irreversible processes on the reversible dynamics. A key challenge for future research is therefore the development of a generalized theory that unifies these two aspects -- reversible dynamics associated to anomalous correlators and irreversible thermalization -- fully accounting for phase-correlations.

\medskip \noindent
{\it Acknowledgement.} 
Funding was provided by Agence Nationale de la Recherche (Grants No. ANR-23-CE30-0021, No. ANR-21-ESRE-0040, No. ANR-21-CE42-0026), by the Marie Sklodowska-Curie Actions project BESCLING (Grant No. 101209943), and by the Italian Ministry of University and Research (MUR) through the Italian Science Fund (FIS2), project SOFT (H53C24001530001). Calculations were performed using HPC resources from Universit\'e C\^ote d’Azur’s Center Azzura, and DNUM-CCUB (Universit\'e de Bourgogne Europe). S.W. acknowledges support from the European Innovation Council, project MULTISCOPE (101185664).

\newpage\quad
\newpage
\baselineskip 11pt

%\begin{widetext}
\onecolumngrid

\section*{ Supplementary information on the article:}

\section{Thermalization: Derivation of a reduced wave turbulence kinetic equation for simulations}

In Fig.~1 (main text), we report numerical simulations of the WT KE~(2) (main text). The one-dimensional interaction allows the KE~(2)  (main text) to be written in a simplified form, which greatly improves the efficiency of its numerical integration. In the following, we derive the simplified version of KE~(2) (main text). 

We start from the coherently coupled NLSE (1) (main text):
\begin{eqnarray}
i\partial_z u_x &=& - \beta \partial_{tt} u_x + \alpha u_x + \gamma (|u_x|^2+ \kappa |u_y|^2)u_x + \gamma \rho u_x^* u_y^2,
\label{eq:nls_x}\\
i\partial_z u_y  &=& - \beta \partial_{tt} u_y - \alpha u_y + \gamma(|u_y|^2+ \kappa |u_x|^2)u_y  + \gamma \rho u_y^* u_x^2.
\label{eq:nls_y}
\end{eqnarray}
The corresponding dispersion relations read $k_{x}(\omega)=\beta \omega^2+\alpha$ and $k_{y}(\omega)=\beta \omega^2-\alpha$. 
The NLSE conserves the power (particle number) $N=N_x+N_y$, with $N_\mu=\frac{1}{T}\int_0^T |u_\mu|^2 dt$ ($\mu=x,y$), where $T$ denotes the size of the numerical window. It also conserves the Hamiltonian  $H = E+U$, which has a linear and a nonlinear contribution. The linear contribution $E=E_k+E_c$ can be split into a kinetic contribution and a resonant coupling contribution:   
\begin{eqnarray}
E_k=\frac{1}{T}  \int_0^T  \beta (|\partial_t u_x|^2+|\partial_t u_y|^2)dt,
\quad
E_c=\frac{1}{T}  \int_0^T  \alpha (|u_x|^2-|u_y|^2)dt.
\label{eq:E_lin}
\end{eqnarray}
The  nonlinear contribution reads 
\begin{eqnarray}
U=\frac{1}{T} \int_0^T \frac{\gamma}{2}\big(|u_x|^4+|u_y|^4\big) + \gamma \kappa |u_x|^2 |u_y|^2 
+\frac{\gamma \rho}{2} \big(u_x^{* 2} u_y^2 + u_x^{2} u_y^{* 2}\big) dt.
\end{eqnarray}
\medskip
Following the standard wave turbulence theory \cite{zakharov92,zakharov04,nazarenko11}, we derive from the vector NLSE (\ref{eq:nls_x}-\ref{eq:nls_y}) the following kinetic equations 
%governing the evolutions of the averaged spectra of the waves $\left<{\tilde \varphi}(\omega_1,z){\tilde \varphi}^*(\omega_2,z)\right>=n_\varphi(\omega,z) \alpha(\omega_1-\omega_2)$ ($\varphi=u_x,u_y$):
\begin{align}
\partial_z n_{x}(\omega,z) &= \kappa^2 {\cal C}oll_i[n_{x},n_{y}] + \rho^2 {\cal C}oll_c[n_{x},n_{y}], 
\\
\partial_z n_{y}(\omega,z) &= \kappa^2 {\cal C}oll_i[n_{y},n_{x}] + \rho^2 {\cal C}oll_c[n_{y},n_{x}],
\end{align}
where $\left< {\tilde u}_\mu(\omega_1,z) {\tilde u}_\mu^*(\omega_2,z) \right>=2\pi n_\mu(\omega_1,z) \delta(\omega_1-\omega_2)$ under the assumption of homogeneous statistics, where ${\tilde u}_\mu(\omega,z)=\int u_\mu(t,z) \exp(-i\omega t) dt$ is the Fourier transform, for $\mu=x,y$.
%$=\int  \varphi(t,z) \exp(-i \omega t) dt$
%We remind that the collision term that comes from the non-resonant self-phase modulation vanishes in 1D.
%where ${\tilde \varphi}(\omega,z)=\frac{1}{\sqrt{2\pi}}\int  \varphi(t,z) \exp(-i \omega t) dt$ is the Fourier transform of $\varphi(t,z)$.
%Note that, according to the standard wave turbulence theory, we have assumed that there is no phase-correlation among the two waves, $\left<u_x(t,z) u_y^*(t,z) \right>=0$, see also \cite{PRX17,onorato20}. 
The collision term for the non-resonant interaction reads:
\begin{eqnarray}
&&\hspace*{-0.2in}
{\cal C}oll_i[n_{x},n_{y}] = \frac{ \gamma^2}{2\pi} \iiint d\omega_{1-3} \,  
 n_{x}(\omega_1) \, n_{y}(\omega_2)\,n_{y}(\omega_3)\,n_{x}(\omega)  \left( n_{x}^{-1}(\omega)+n_{y}^{-1}(\omega_3)-n_{y}^{-1}(\omega_2)-n_{x}^{-1}(\omega_1)\right) \nonumber\\ 
&&\hspace*{2.3in}
\times \, \delta\big(k_x(\omega_1)+k_y(\omega_2)-k_y(\omega_3)-k_x(\omega)\big) \
\delta(\omega_1+\omega_2-\omega_3-\omega), 
\label{eq:wtcollu}\\
&&\hspace*{-0.2in}
{\cal C}oll_i[n_{y},n_{x}] = \frac{ \gamma^2}{2\pi} \iiint d\omega_{1-3} \,  
 n_{y}(\omega_1) \, n_{x}(\omega_2)\,n_{x}(\omega_3)\,n_{y}(\omega)  \left( n_{y}^{-1}(\omega)+n_{x}^{-1}(\omega_3)-n_{x}^{-1}(\omega_2)-n_{y}^{-1}(\omega_1)\right) \nonumber \\ 
&&\hspace*{2.3in}
\times \, \delta\big(k_y(\omega_1)+k_x(\omega_2)-k_x(\omega_3)-k_y(\omega)\big) \
\delta(\omega_1+\omega_2-\omega_3-\omega),
\label{eq:wtcollv}
\end{eqnarray}
with $d\omega_{1-3}=d\omega_1 d\omega_2 d\omega_3$.
The collision term for the resonant interaction reads:
\begin{eqnarray}
&&\hspace*{-0.2in}
{\cal C}oll_c[n_{x},n_{y}] = \frac{ 2\gamma^2}{2\pi} \iiint d\omega_{1-3} \,  
 n_{y}(\omega_1) \, n_{y}(\omega_2)\,n_{x}(\omega_3)\,n_{x}(\omega)  \left( n_{x}^{-1}(\omega)+n_{x}^{-1}(\omega_3)-n_{y}^{-1}(\omega_2)-n_{y}^{-1}(\omega_1)\right) \nonumber\\ 
&&\hspace*{2.3in}
\times \, \delta\big(k_y(\omega_1)+k_y(\omega_2)-k_x(\omega_3)-k_x(\omega)\big) \
\delta(\omega_1+\omega_2-\omega_3-\omega), 
\label{eq:wtcollu}\\
&&\hspace*{-0.2in}
{\cal C}oll_c[n_{y},n_{x}] = \frac{ 2\gamma^2}{2\pi} \iiint d\omega_{1-3} \,  
 n_{x}(\omega_1) \, n_{x}(\omega_2)\,n_{y}(\omega_3)\,n_{y}(\omega)  \left( n_{y}^{-1}(\omega)+n_{y}^{-1}(\omega_3)-n_{x}^{-1}(\omega_2)-n_{x}^{-1}(\omega_1)\right) \nonumber \\ 
&&\hspace*{2.3in}
\times \, \delta\big(k_x(\omega_1)+k_x(\omega_2)-k_y(\omega_3)-k_y(\omega)\big) \
\delta(\omega_1+\omega_2-\omega_3-\omega).
\label{eq:wtcollv}
\end{eqnarray}
Note that, because of the degenerate resonances inherent to the purely one-dimensional problem considered here, the collision term describing the self-interaction (first nonlinear term in Eqs.(\ref{eq:nls_x}-\ref{eq:nls_y})) vanishes identically, see \cite{zakharov92,zakharov04,nazarenko11}.
We focus on the analysis of the resonant collision term.
Two integrals can be computed owing to the Dirac $\delta-$functions, which gives:
\begin{eqnarray}
&&\hspace*{-0.2in}
\partial_z n_{x}(\omega,z) = \kappa^2 {\cal C}oll_i[n_{x},n_{y}] + \frac{2 \rho^2 \gamma^2}{4\pi |\beta|} \int   
 n_{y}(\omega +
 {\omega'}) \, n_{y}(\omega-\nu/{\omega'})\,n_{x}(\omega +{\omega'}-\nu/{\omega'})\,n_{x}(\omega) \nonumber\\ 
&&\hspace*{2.2in}
 \times \left( n_{x}^{-1}(\omega)+n_{x}^{-1}(\omega  + {\omega'}-\nu/{\omega'})-n_{y}^{-1}(\omega +{\omega'})-n_{y}^{-1}(\omega-\nu/{\omega'})\right) \frac{d{\omega'}}{|{\omega'}|} , \\ 
&&\hspace*{-0.2in}
\partial_z n_{y}(\omega,z) = \kappa^2 {\cal C}oll_i[n_{y},n_{x}] +\frac{2 \rho^2 \gamma^2}{4\pi |\beta|} \int   
 n_{x}(\omega +{\omega'}) \, n_{x}(\omega + \nu/{\omega'})\,n_{y}(\omega +{\omega'} +
 \nu/{\omega'})\,n_{y}(\omega) \nonumber\\ 
&&\hspace*{2.2in}
 \times  \left( n_{y}^{-1}(\omega)+n_{y}^{-1}(\omega +{\omega'} +\nu/{\omega'})-n_{x}^{-1}(\omega +{\omega'})-n_{x}^{-1}(\omega +\nu/{\omega'})\right) \frac{d{\omega'}}{|{\omega'}|} ,
\end{eqnarray}
where $\nu = 2 \alpha/\beta$. 
\begin{eqnarray}
&&\hspace*{-0.2in}
\partial_z n_{x}(\omega,z) = \frac{\kappa^2 \gamma^2}{4\pi |\beta|} \int  
 n_{x}({\omega'}) \, n_{y}(\omega)\,n_{y}({\omega'})\,n_{x}(\omega)  \left( n_{x}^{-1}(\omega)+n_{y}^{-1}({\omega'})-n_{y}^{-1}(\omega)-n_{x}^{-1}({\omega'})\right)  \frac{d{\omega'}}{|{\omega'}-\omega|} \nonumber\\
&&\hspace*{0.3in}+ \frac{2 \rho^2 \gamma^2}{4\pi |\beta|} \int   
 n_{y}(\omega + \sqrt{|\nu|} {\eta}) 
 \, n_{y}(\omega-\sqrt{|\nu|}\sigma /{\eta} )\,n_{x}(\omega + \sqrt{|\nu|} {\eta}-\sqrt{|\nu|} \sigma /{\eta} )\,n_{x}(\omega) \nonumber\\ 
&&\hspace*{0.3in}
\times \left( n_{x}^{-1}(\omega)+n_{x}^{-1}(\omega +\sqrt{|\nu|} {\eta}- \sqrt{|\nu|} \sigma / {\eta})-n_{y}^{-1} (\omega +\sqrt{|\nu|} {\eta})-n_{y}^{-1}(\omega-\sqrt{|\nu|} \sigma / {\eta})\right) \frac{d{\eta}}{|{\eta}|} , \\ 
&&\hspace*{-0.2in}
\partial_z n_{y}(\omega,z) = \frac{\kappa^2 \gamma^2}{4\pi |\beta|} \int  
 n_{x}({\omega'}) \, n_{y}(\omega)\,n_{y}({\omega'})\,n_{x}(\omega)  \left( n_{x}^{-1}(\omega)+n_{y}^{-1}({\omega'})-n_{y}^{-1}(\omega)-n_{x}^{-1}({\omega'})\right)  \frac{d{\omega'}}{|{\omega'}-\omega|} \nonumber\\
&&\hspace*{0.3in}+\frac{2 \rho^2 \gamma^2}{4\pi |\beta|} \int   
 n_{x}(\omega + \sqrt{|\nu|} {\eta}) \, n_{x}(\omega + \sqrt{|\nu|} \sigma/ {\eta})\,n_{y}(\omega + \sqrt{|\nu|} {\eta} +
\sqrt{|\nu|} \sigma /{\eta} )\,n_{y}(\omega) \nonumber\\ 
&&\hspace*{0.3in}
\times  \left( n_{y}^{-1}(\omega)+n_{y}^{-1}(\omega +\sqrt{|\nu|} {\eta} +\sqrt{|\nu|} \sigma/ {\eta})
 -n_{x}^{-1}(\omega + \sqrt{|\nu|} {\eta})-n_{x}^{-1}(\omega +\sqrt{|\nu|} \sigma /{\eta} )\right) \frac{d{\eta}}{|{\eta}|} ,
\end{eqnarray}
with $\sigma={\rm sgn}(\nu)$.
%It can be checked that $\partial_z(N_U+N_V)=0$.

The singularity in the resonant collision terms (proportional to $\rho^2$) can be rewritten in a convenient form by the change of variable $\eta \to s e^\theta$, $s=\pm 1$, so that the kinetic equations take the form 
%\begin{widetext}
\begin{eqnarray}
&&\hspace*{-0.2in}
\partial_z n_{x}(\omega,z) = 
%{\cal C}oll_X[n_U,n_V] \nonumber\\ 
\frac{\kappa^2 \gamma^2}{4\pi |\beta|} \int  
 n_{x}({\omega'}) \, n_{y}(\omega)\,n_{y}({\omega'})\,n_{x}(\omega)  \left( n_{x}^{-1}(\omega)+n_{y}^{-1}({\omega'})-n_{y}^{-1}(\omega)-n_{x}^{-1}({\omega'})\right)  \frac{d{\omega'}}{|{\omega'}-\omega|} \nonumber\\
&&\hspace*{0.3in}
+ \frac{2\rho^2 \gamma^2}{4\pi |\beta|} \sum_s 
\int   
 n_{y}(\omega + \sqrt{|\nu|} s e^\theta) 
 \, n_{y}(\omega-\sqrt{|\nu|}\sigma s e^{-\theta} )\,n_{x}(\omega + \sqrt{|\nu|} s e^\theta-\sqrt{|\nu|} \sigma s e^{-\theta} )\,n_{x}(\omega) \nonumber\\ 
&&\hspace*{0.3in}
\times \left( n_{x}^{-1}(\omega)+n_{x}^{-1}(\omega +\sqrt{|\nu|} s e^\theta - \sqrt{|\nu|} \sigma s  e^{-\theta})-n_{y}^{-1}(\omega +\sqrt{|\nu|} s  e^\theta )-n_{y}^{-1}(\omega-\sqrt{|\nu|} \sigma s  e^{-\theta})\right) d\theta  ,
\label{eq:red_wtkin_{x}} 
\\ 
&&\hspace*{-0.2in}
\partial_z n_{y}(\omega,z) = 
%{\cal C}oll_X[n_V,n_U] \nonumber\\ 
\frac{\kappa^2 \gamma^2}{4\pi |\beta|} \int   
 n_{y}({\omega'}) \, n_{x}(\omega)\,n_{x}({\omega'})\,n_{y}(\omega)  \left( n_{y}^{-1}(\omega)+n_{x}^{-1}({\omega'})-n_{x}^{-1}(\omega)-n_{y}^{-1}({\omega'})\right) \frac{d{\omega'}}{|{\omega'}-\omega|} \nonumber\\ 
&&\hspace*{0.3in}
+\frac{2 \rho^2 \gamma^2}{4\pi |\beta|} \sum_s  \int   
 n_{x}(\omega + \sqrt{|\nu|} s e^{\theta}  ) \, n_{x}(\omega + \sqrt{|\nu|} \sigma s e^{-\theta} )\,n_{y}(\omega + \sqrt{|\nu|} s e^{\theta}  +
\sqrt{|\nu|} \sigma s e^{-\theta}  )\,n_{y}(\omega) 
\nonumber\\ 
&&\hspace*{0.3in}
\times \left( n_{y}^{-1}(\omega)+n_{y}^{-1}(\omega +\sqrt{|\nu|} s  e^{\theta}+\sqrt{|\nu|} \sigma s e^{-\theta} )
 -n_{x}^{-1}(\omega + \sqrt{|\nu|} s e^{\theta} )-n_{x}^{-1}(\omega +\sqrt{|\nu|} \sigma s e^{-\theta}  )\right) d\theta.
\label{eq:red_wtkin_{y}} 
 \end{eqnarray}
The kinetic equations conserve the total power (density) $N=\frac{1}{2\pi}\sum_\mu \int n_{\mu}(\omega,z) d\omega$, with $\mu=x,y$. Note that if $\sigma=1$, then
\begin{eqnarray}
&&\hspace*{-0.2in}
\partial_z N_{x}(z) = \frac{2\rho^2 \gamma^2}{2\pi |\beta|}  \int d\omega  \int d\theta 
 n_{y}(\omega+\sqrt{|\nu|}  \cosh \theta) \, n_{y}(\omega -\sqrt{|\nu|}  \cosh \theta)\,n_{x}(\omega +\sqrt{|\nu|}  \sinh \theta)\,n_{x}(\omega-\sqrt{|\nu|}  \sinh \theta) \nonumber\\ 
&&\hspace*{0.3in}
\times \left( n_{x}^{-1} (\omega +\sqrt{|\nu|}  \sinh \theta)+n_{x}^{-1} (\omega -\sqrt{|\nu|}  \sinh \theta)-n_{y}^{-1}(\omega +\sqrt{|\nu|}  \cosh \theta)-n_{y}^{-1} (\omega - \sqrt{|\nu|}  \cosh \theta)\right) ,
\label{eq:dzNx} \\ 
&&\hspace*{-0.2in}
\partial_z N_{y}(z) =  - \partial_z N_{x}(z)   .
\label{eq:dzNy}
\end{eqnarray}
If $\sigma=-1$, the same result is obtained by changing $\cosh(\theta)$ and $\sinh(\theta)$. The kinetic equation also conserves the linear energy $E=\frac{1}{2\pi}\sum_\mu \int k_{\mu}(\omega)n_{\mu}(\omega,z) d\omega$, with $\mu=x,y$.

\newpage

\begin{figure}
    \centering
    \includegraphics[width=0.9\linewidth]{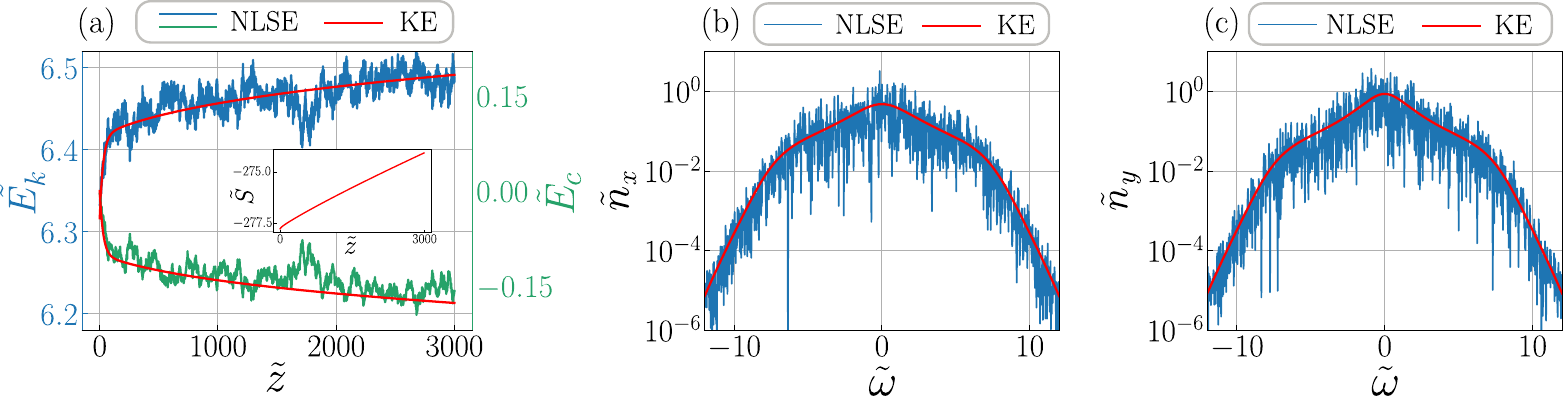}
    \caption{
    \baselineskip 10pt
    {\bf Irreversible thermalization.} Numerical simulations of the NLSE (\ref{eq:nls_x}-\ref{eq:nls_y}) \blue{(blue and green  lines)}, and \blue{numerical simulations of} the wave turbulence KE (\ref{eq:red_wtkin_{x}}-\ref{eq:red_wtkin_{y}}) (red lines): (a) Evolution of the \blue{normalized kinetic contribution to the energy ${\tilde E}_k({\tilde z})$ (blue line)}, and \blue{the normalized coherent coupling contribution ${\tilde E}_c({\tilde z})$} (green line). The inset \blue{in (a) shows the corresponding evolution of the normalized entropy ${\tilde S}({\tilde z})$ obtained from the simulation of the KE (\ref{eq:red_wtkin_{x}}-\ref{eq:red_wtkin_{y}}) (red line)}. Spectra of the waves \blue{${\tilde n}_x({\tilde \omega})$ (b), and ${\tilde n}_y({\tilde \omega})$ (c) at ${\tilde z}=3000$, from NLSE (blue lines), and KE (red lines), simulations}. 
    The thermalization process is characterized by a growth of the kinetic energy \blue{${\tilde E}_k({\tilde z})$}, which entails a decrease of \blue{${\tilde E}_c({\tilde z})$, because the total linear energy is conserved ${\tilde E}={\tilde E}_k({\tilde z})+{\tilde E}_c({\tilde z}) \simeq$~const, in the considered weakly nonlinear regime $|{\tilde E}/{\tilde U}| \gg 1$}. Recalling that \blue{${\tilde E}_c({\tilde z}) = {\tilde \alpha} ({\tilde N}_x-{\tilde N}_y)$}, thermalization is characterized by a power transfer from the $x$-to-$y$ axis, which entails  the increase of the degree of polarization ${\cal P}$, see Fig.~1 (main text). 
    \blue{In panels (a)-(c), variables marked with a tilde ($\sim$) denote normalized quantities, as defined in the text.} 
    An average over 5 realizations of the initial condition (i.e., initial random spectral phases) has been considered for the NLSE simulations. \blue{Parameters are the same as in Fig.~1 (main text): ${\tilde \alpha}=1, {\tilde N}_x^0={\tilde N}_y^0=0.5, {\tilde \sigma}=0.8 \pi, \kappa=2/3$, $|{\tilde E}/{\tilde U}| \simeq {\tilde \sigma}^2$.}
    }
    \label{fig:thermal}
\end{figure}

The reduced form of the wave turbulence KE (\ref{eq:red_wtkin_{x}}-\ref{eq:red_wtkin_{y}}) has been numerically integrated. The results have been compared with the simulations of the NLSE (\ref{eq:nls_x}-\ref{eq:nls_y}), and a good agreement is obtained without using adjustable parameters, see Fig.~1 (main text), and Fig.~\ref{fig:thermal}. 
\blue{
For the numerical simulations, we have normalized the NLSE (\ref{eq:nls_x}-\ref{eq:nls_y}) and the wave turbulence KE (\ref{eq:red_wtkin_{x}}-\ref{eq:red_wtkin_{y}}). The dimensionless variables are ${\tilde z}=z/L_{nl}$, ${\tilde t}=t/\tau_0$, ${\tilde u}_\mu=u_\mu/\sqrt{N}$, with $\mu=x,y$, and $L_{nl}=1/(\gamma N)$ the nonlinear length, $\tau_0=\sqrt{|\beta| L_{nl}}$ the `healing time'. With this normalization, we have ${\tilde N}_\mu=N_\mu/N$, ${\tilde E}_k=E_k L_{nl}/N$, ${\tilde E}_c=E_c L_{nl}/N$, ${\tilde U}=U L_{nl}/N$, ${\tilde n}_\mu={n}_\mu/(N \tau_0)$, ${\tilde \omega}=\omega \tau_0$, ${\tilde S}({\tilde z})=\frac{1}{2\pi} \int \log({\tilde n}({\tilde \omega},{\tilde z})) d{\tilde \omega}$, ${\tilde \sigma}=\sigma \tau_0$, and ${\tilde \alpha}=\alpha L_{nl}$.
The corresponding evolution during propagation of the energies ${\tilde E}_k({\tilde z})$ and  ${\tilde E}_c({\tilde z})$ are reported in Fig.~\ref{fig:thermal}(a).}
%Repolarization is primarily driven by the thermalization process, which reflects the natural tendency of waves to increase disorder. 
%This manifests itself by a spectral broadening of the waves in Fig.~1. Since the kinetic energy is related to the spectral width of the waves, 
%spectral broadening results in an increase in $E_k(z)$ as the waves propagate, see Fig.~2.
We remark that, unlike conventional nonlinear systems where the kinetic energy \blue{${\tilde E}_k$} is the sole contribution to the linear energy \blue{${\tilde E}$}, our system includes an additional contribution arising from the coherent coupling, \blue{${\tilde E}={\tilde E}_k+{\tilde E}_c$, which is conserved  ${\tilde E} \simeq$~const in the weakly nonlinear regime $|{\tilde E}/{\tilde U}| \gg 1$}. Consequently, the growth in the kinetic energy \blue{${\tilde E}_k$} is no longer constrained, since it can be offset by a corresponding decrease in \blue{${\tilde E}_c$}, as illustrated in Fig.~\ref{fig:thermal}(a) \cite{OE08}. \blue{Recalling that ${\tilde E}_c(z) = {\tilde \alpha} ({\tilde N}_x-{\tilde N}_y)$ [see Eq.(\ref{eq:E_lin})], thermalization manifests itself by an irreversible transfer of power from the $x-$axis to the $y-$axis, thus leading to the emergence of a nonvanishing degree of polarization, as illustrated in Fig.~1(a)-(b) in the main text.}

\section{Anomalous correlator kinetic equation}

\subsection{Derivation of Eqs.(4-5) (main text)} 

We start from the NLSE (\ref{eq:nls_x}-\ref{eq:nls_y}) and define the normal correlator 
$U_\mu(t_1,t_2,z)=\left< u_\mu(t_1,z)u_\mu(t_2,z)^* \right>$ for $\mu=x,y$, and the anomalous correlator describing a correlation among the waves $u_x$ and $u_y$: $W(t_1,t_2,z)=\left< u_y(t_1,z)u_x^*(t_2,z) \right>$.
Following the usual procedure, we derive the equations for the correlators $U_\mu$ and $W$, which involve fourth-order moments of the waves that can be expanded into products of second-order moments in the weakly nonlinear regime of Gaussian statistics. Here, caution should be exercized in keeping all anomalous correlations between the waves. We obtain
\begin{align}
i\partial_z U_x^{12} =& - \beta(\partial_{t_1}^2-\partial_{t_2}^2)U_x^{12}
+ \gamma U_x^{12}[2(U_x^{11}-U_x^{22})+\kappa(U_y^{11}-U_y^{22})]
+ \gamma \kappa \big[ W^{12}(W^{11}+W^{11*}) - W^{21*}(W^{22}+W^{22*}) \big],
\nonumber \\
i\partial_z U_y^{12} =& - \beta(\partial_{t_1}^2-\partial_{t_2}^2)U_y^{12}
+\gamma U_y^{12}[2(U_y^{11}-U_y^{22})+\kappa(U_x^{11}-U_x^{22})]
+ \gamma \kappa \big[ W^{21*}(W^{11}+W^{11*}) - W^{12}(W^{22}+W^{22*}) \big],
\nonumber \\
i\partial_z W^{12} =& - \beta(\partial_{t_1}^2-\partial_{t_2}^2)W^{12}
-2\alpha W^{12}
+\gamma W^{12} \big[2(U_y^{11}-U_x^{22})+\kappa(U_x^{11}-U_y^{22})\big]
\nonumber \\
&+\gamma \kappa \big[U_x^{12}(W^{11}+W^{11*})-U_y^{12}(W^{22}+W^{22*}) \big]  ,
\nonumber 
\end{align}
where  $U_\mu^{ij}=U_\mu(t_i,t_j,z)$ for $\mu=x,y$, and $W^{ij}=W(t_i,t_j,z)$.
Introducing the variables $t=(t_1+t_2)/2, \tau=t_1-t_2$, this also reads
\begin{align}
i\partial_z U_x =& - 2 \beta \partial_{t \tau} U_x + \gamma {\cal L}^{U_x} ,\\
i\partial_z U_y =& - 2 \beta \partial_{t \tau} U_y + \gamma {\cal L}^{U_y} ,\\
i\partial_z W =& - 2 \beta\partial_{t \tau}  W
-2\alpha W + \gamma {\cal L}^{W} ,
\end{align}
with the three nonlinear terms
\begin{align}
{\cal L}^{U_x} (t,\tau) =& 
U_x(t,\tau) [ (2U_x+\kappa U_y)(t+\tau/2,0)-(2U_x+\kappa U_y)(t-\tau/2,0)] 
\nonumber
\\
&
+ 2 \kappa \big[ W(t,\tau) W^r(t+\tau/2,0) - W(t,-\tau)^* W^r(t-\tau/2,0)  \big],\\
{\cal L}^{U_y} (t,\tau) =& 
U_y(t,\tau) [ (2U_y+\kappa U_x)(t+\tau/2,0)-(2U_y+\kappa U_x)(t-\tau/2,0)]
\nonumber
\\
&
+ 2 \kappa \big[ W(t,-\tau)^* W^r(t+\tau/2,0) - W(t,\tau) W^r(t-\tau/2,0)  \big], \\
{\cal L}^W(t,\tau)=&  W(t,\tau) \big[(2U_y+\kappa U_x)(t+\tau/2,0)-(2U_x+\kappa U_y)(t-\tau/2,0)\big]
\nonumber
\\
&
+2 \kappa \big[U_x(t,\tau) W^r(t+\tau/2) -U_y (t,\tau) W^r(t-\tau/2,0) \big]  ,
\end{align}
where $W^r={\rm Re}(W)$.
We can make use of the assumption of quasi-stationary statistics, that is to say the time correlation $t_c$ (i.e. the microscopic time scale of random fluctuations) is much smaller than the macroscopic time scale $t_s$ of stationary (homogeneous) statistics. In the regime $\varepsilon = t_c/t_s \ll 1$, the three nonlinear terms can be expanded up to first order as
\begin{align}
{\cal L}^{U_x} (t,\tau) =& 
U_x(t,\tau)  \tau \partial_t (2U_x+\kappa U_y)(t,0) 
+ 2 \kappa  W^r(t,0)  \big[ W(t,\tau) - W(t,-\tau)^* \big]
+ \kappa \tau \partial_t W^r(t,0) \big[ W(t,\tau) + W(t,-\tau)^* \big] ,\\
{\cal L}^{U_y} (t,\tau) =& 
U_y(t,\tau) \tau \partial_t  (2U_y+\kappa U_x)(t,0) 
- 2 \kappa W^r(t,0)  \big[  W(t,\tau) - W(t,-\tau)^*   \big]
+  \kappa \tau \partial_t W^r(t,0)  \big[ W(t,\tau) +W(t,-\tau)^*  \big]
, \\
\nonumber
{\cal L}^W(t,\tau)=& (2-\kappa) W(t,\tau) ( U_y -U_x)(t,0)
+(2+\kappa)W(t,\tau) \frac{\tau}{2}\partial_t (U_y+U_x)(t,0) \\
&
+ 2\kappa W^r(t,0) (U_x - U_y)(t,\tau)  
 +\kappa \tau \partial_t W^r(t,0) (U_x+U_y)(t,\tau) .
\end{align}
This gives
\begin{align}
i\partial_z U_x(t,\tau,z) =& - 2 \beta\partial_{t \tau} U_x +  
\gamma\tau U_x \partial_t \big( 2 N_x(t) + \kappa N_y(t) \big)
-2\gamma\kappa W_0^r(t) \big( W^*(t,-\tau) - W(t,\tau) \big)\nonumber
\\
&
+ \gamma \kappa \tau \partial_t W_0^r(t) \big[ W(t,\tau) + W(t,-\tau)^* \big]  ,
\label{eq:U_x}  \\
i\partial_z U_y(t,\tau,z) =& - 2 \beta \partial_{t \tau} U_y +  
\gamma\tau U_y \partial_t \big( 2 N_y(t) + \kappa N_x(t) \big)
-2\gamma\kappa W_0^r(t) \big( W(t,\tau)-W^*(t,-\tau) \big) 
\nonumber
\\
& + \gamma \kappa \tau \partial_t W_0^r(t) \big[ W(t,\tau) + W(t,-\tau)^* \big] ,
\label{eq:U_y}  \\
i\partial_z W(t,\tau,z) =& - 2\beta \partial_{t \tau} W  
+W\Big( \gamma(2-\kappa)\big(N_y(t)-N_x(t)\big)-2\alpha   \Big) 
- 2\gamma\kappa \big( U_y-U_x \big) W_0^r(t) \nonumber \\
&+\gamma(1+\kappa/2)\tau W \partial_t \big( N_y(t)+N_x(t) \big)
+\gamma\kappa \tau \big( U_y+U_x \big) \partial_t W_0^r(t)  , 
\label{eq:W} 
\end{align}
where $W_0(t,z)=W(t,\tau=0,z)$, and $N_\mu(t,z)=U_\mu(t,\tau=0,z)$.

The equations for the normal correlator $n_\mu(t,\omega,z)=\int U_\mu(t,\tau,z) \exp(-i\omega \tau) d\tau$, and the anomalous correlator $m(t,\omega,z)=\int W(t,\tau,z) \exp(-i\omega \tau) d\tau$, are obtained by taking the Fourier transform of Eqs.(\ref{eq:U_x}-\ref{eq:W}), leading to the set of Eqs.(4-5) (main text):
\begin{align}
\partial_z n_x(\omega,t) =& - 2\beta \omega \partial_t n_x 
+ \gamma \partial_t \big(2 N_x(t) + \kappa N_y(t) \big)\partial_\omega n_x
+2\gamma\kappa \partial_t M^r(t) \partial_\omega m^r(\omega,t)
+4 \gamma\kappa M^r(t) m^i(\omega,t) ,
\label{eq:nx_vl_supp} \\
\partial_z n_y(\omega,t) =& - 2 \beta\omega \partial_t n_y + \gamma \partial_t \big(2 N_y(t) + \kappa N_x(t) \big)\partial_\omega n_y
+2\gamma\kappa \partial_t M^r(t) \partial_\omega m^r(\omega,t)
-4 \gamma\kappa M^r(t) m^i(\omega,t)  ,
\label{eq:ny_vl_supp} \\
\partial_z m(\omega,t) =& -2\beta \omega \partial_t m 
+\gamma(1+\kappa/2) \partial_t \big( N_x(t)+N_y(t) \big) \partial_\omega m
+\gamma\kappa \partial_t M^r(t)\partial_\omega \big( n_x+n_y \big)
\nonumber \\
&-i m \Big( \gamma(2-\kappa)\big(N_y(t)-N_x(t)\big)-2\alpha   \Big)  
- 2i\gamma\kappa \big( n_x - n_y \big) M^r(t)   ,
\label{eq:m_vl_supp}
\end{align}
where $M(t,z)=W_0(t,z)=\frac{1}{2\pi}\int m(t,\omega,z)d\omega$, and ${N}_{\mu}(t,z)=\frac{1}{2\pi}\int n_{\mu}(t,\omega,z)d\omega$, while $m^{r,i}$ ($M^{r,i}$) denote the real and imaginary parts of $m$ ($M$). 

Note that the system of Eqs.(\ref{eq:nx_vl_supp}-\ref{eq:m_vl_supp}) is formally reversible in the `time' variable $z$, i.e., it is invariant under the transformation: $(z \to -z, \omega \to -\omega, m \to m^*)$.  

\subsection{Stability analysis for the emergence of phase-correlations: Derivation of Eq.(6) (main text)} 

We assume that the initial random waves $u_\mu(t,z=0)$ have a homogeneous statistics with spectra $n_\mu^0(\omega)$, and are uncorrelated with each other, so that the initial anomalous correlator is zero, $m(\omega,t,z=0)=0$. We perform a linear stability analysis of Eqs.(\ref{eq:nx_vl_supp}-\ref{eq:m_vl_supp}) around this solution with 
\begin{align}
n_\mu(t,\omega,z) &= n_\mu^0(\omega) + \delta n_\mu(t,\omega,z), \quad \mu=x,y ,
\label{eq:pert_n}\\
m(t,\omega,z) &= \delta m(t,\omega,z),
\label{eq:pert_m}
\end{align}
with $\delta n_\mu(t,\omega,z) \ll n_\mu^0(\omega)$ and $|\delta m(t,\omega,z)| \ll n_\mu^0(\omega)$. 
Introducing (\ref{eq:pert_n}-\ref{eq:pert_m}) into Eqs.(\ref{eq:nx_vl_supp}-\ref{eq:m_vl_supp}), we obtain after linearization 
\begin{align}
\partial_z \delta n_x(t,\omega) =& - 2\beta \omega \partial_t \delta n_x + \gamma\partial_\omega n_x^0 \partial_t \big(2 \delta N_x(t) + \kappa \delta N_y(t) \big),
\label{eq:deltanx_vl_supp} \\
\partial_z \delta n_y(t,\omega) =& - 2 \beta\omega \partial_t \delta n_y + \gamma\partial_\omega n_y^0 \partial_t \big(2 \delta N_y(t) + \kappa \delta N_x(t) \big),
\label{eq:deltany_vl_supp} \\
\partial_z \delta m(t,\omega) =& -2\beta \omega \partial_t \delta m 
-i \delta m \Big( \gamma(2-\kappa)\big(N_y^0-N_x^0\big)-2\alpha   \Big)  
- 2i\gamma\kappa \big( n_x^0 - n_y^0 \big) \delta M^r(t) \nonumber \\
& +\gamma\kappa\partial_\omega \big( n_x^0+n_y^0 \big) \partial_t \delta M^r(t),
\label{eq:deltam_vl_supp}
\end{align}
where $N_\mu^0 = \frac{1}{2\pi}\int n_\mu^0(\omega) d\omega$, $\delta N_\mu(t,z) = \frac{1}{2\pi}\int \delta n_\mu(t,\omega,z) d\omega$, $\delta M(t,z) = \frac{1}{2\pi}\int \delta m(t,\omega,z) d\omega$. Note that the evolution of $\delta n_\mu$ in (\ref{eq:deltanx_vl_supp}-\ref{eq:deltany_vl_supp}) is decoupled from the evolution $\delta m$ in (\ref{eq:deltam_vl_supp}). In the following we study separately the stability analysis for the anomalous correlator, and the normal correlators.

\subsubsection{Stability analysis of the anomalous correlator, Eq.(\ref{eq:deltam_vl_supp})}

We perform the stability analysis of the anomalous correlator by  introducing $\delta {\tilde m}(\Omega,\omega,z)=\int \delta m(t,\omega,z)\exp(-i\Omega t) dt$, we obtain from Eq.(\ref{eq:m_vl_supp}) 
\begin{align*}
i\partial_z \delta {\tilde m}(\Omega,\omega,z) =& 
\big( 2\beta \omega \Omega + \gamma (2-\kappa)(N_y^0-N_x^0) -2 \alpha  \big)\delta {\tilde m}
\\
&+ \frac{\gamma \kappa}{2\pi} 
\big(n_x^0(\omega)- n_y^0(\omega)-\frac{\Omega}{2}\partial_\omega n_y^{0}(\omega)-\frac{\Omega}{2}\partial_\omega n_x^{0}(\omega) \big) 
 \int \big(  \delta {\tilde m}(\Omega,\omega_1,z) +  \delta {\tilde m}^*(-\Omega,\omega_1,z) \big) d\omega_1.
\end{align*}
To derive the dispersion relation, it proves convenient to introduce the function $\delta {\tilde p}(\Omega,\omega,z)=\delta {\tilde m}^*(\Omega,-\omega,z)$, which gives the system
\begin{align*}
i\partial_z \delta {\tilde m}(\Omega,\omega,z) =& 
\big( 2\beta \omega \Omega + \gamma (2-\kappa)(N_y^0-N_x^0) -2 \alpha  \big)\delta {\tilde m}
\\
&+ \frac{\gamma \kappa}{2\pi} 
\big(n_x^0(\omega)- n_y^0(\omega)-\frac{\Omega}{2}\partial_\omega n_y^0(\omega)-\frac{\Omega}{2}\partial_\omega n_x^0(\omega) \big) 
 \int \big(  \delta {\tilde m}(\Omega,\omega_1,z) +  \delta {\tilde p}(\Omega,\omega_1,z) \big) d\omega_1   , \\
-i\partial_z \delta {\tilde p}(\Omega,\omega,z) =& 
\big( -2\beta \omega \Omega + \gamma (2-\kappa)(N_y^0-N_x^0) -2 \alpha  \big)\delta {\tilde m}
\\
&+ \frac{\gamma \kappa}{2\pi} 
\big(n_y^0(\omega)- n_x^0(\omega)+\frac{\Omega}{2}\partial_\omega n_y^0(\omega)+\frac{\Omega}{2}\partial_\omega n_x^0(\omega) \big) 
 \int \big(  \delta {\tilde m}(\Omega,\omega_1,z) +  \delta {\tilde p}(\Omega,\omega_1,z) \big) d\omega_1   .
\end{align*}
Introducing the Laplace transforms of the anomalous correlator  
$\delta {\hat m}(\Omega,\omega,\lambda)=\int_0^z ds \delta {\tilde m}(\Omega,\omega,s) \exp(-\lambda s)$, $\delta {\hat p}(\Omega,\omega,\lambda)=\int_0^z ds \delta {\tilde p}(\Omega,\omega,s) \exp(-\lambda s)$, we obtain the dispersion relation for the instability growth-rate of the anomalous correlator ${\rm Re}[\lambda(\Omega)]$:
\begin{eqnarray}
\frac{2\pi}{\gamma\kappa} = \int  
\frac{n_x^0(\omega)-n_y^0(\omega)-(\Omega/2)\partial_\omega \big(n_x^0(\omega)+n_y^0(\omega)\big)}{i \lambda-2\beta\omega \Omega - \gamma(2-\kappa)(N_y^0-N_x^0)+2\alpha} 
+\frac{n_x^0(\omega)-n_y^0(\omega)+(\Omega/2)\partial_\omega \big(n_x^0(\omega)+n_y^0(\omega)\big)}{-i \lambda+2\beta\omega \Omega - \gamma(2-\kappa)(N_y^0-N_x^0)+2\alpha} 
d\omega.
\label{eq:disp_rel_supp}
\end{eqnarray}
We have considered initial Gaussian spectra of the form: $n_\mu^0(\omega)=\frac{\sqrt{2\pi}}{\sigma} N_\mu^0 \exp(-\omega^2/(2 \sigma^2))$. The analysis shows that, in general, ${\rm Re}[\lambda(\Omega)]$ is peaked on $\Omega=0$, i.e., the homogeneous mode is the most unstable. Note in this respect that the growth-rate of the homogeneous mode does not depend on the initial spectrum: Regardless of the particular form of $n_\mu^0(\omega)$, the growth-rate for the homogeneous mode reads:
\begin{equation}
\lambda(\Omega=0)=  
 \sqrt{\big(\gamma(2-\kappa)\Delta N^0-2\alpha\big)\big( \gamma(3\kappa-2)\Delta N^0+2\alpha\big)}.    
\end{equation}
Note that this expression recovers Eq.(6) (main text) for $\kappa=2/3$.

\begin{figure}[!h]
    \centering
    \includegraphics[width=0.95\linewidth]{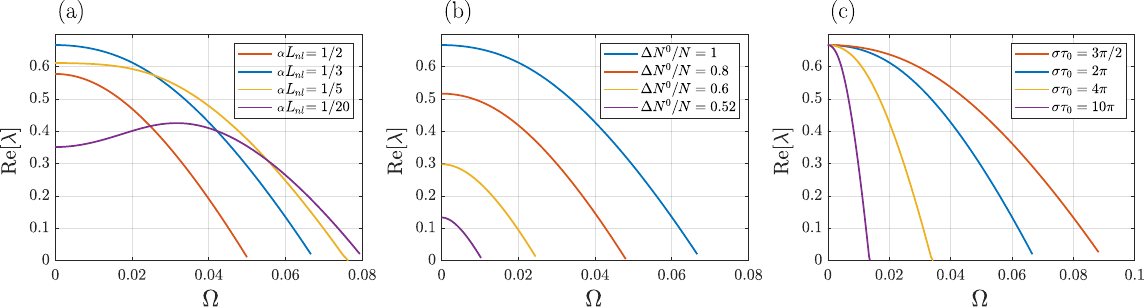}
    \caption{
    \baselineskip 10pt
    \textbf{Impact of the various parameters on $\rm{Re}[\lambda]$.} 
    %In all panels, we have $\sigma \tau_0= \pi\sigma_f$. 
    Panel (a) shows the influence of varying $\alpha L_{nl}$ while keeping $\Delta N^0/N=1$ and $\sigma \tau_0 = 2 \pi$. Panel (b) shows the influence of varying $\Delta N^0/N$ while keeping $\alpha L_{nl}=1/3$ and $\sigma \tau_0 = 2 \pi $. Panel (c) shows the influence of varying $\sigma \tau_0$ while keeping $\alpha L_{nl}=1/3$ and $\Delta N^0 /N= 1$.}
    \label{fig:stability}
\end{figure}

Fig.~\ref{fig:stability} shows the influence of the various parameters on $\rm{Re}[\lambda]$. We can see that in most cases, the maximum growth, i.e. the maximum of $\rm{Re}[\lambda]$, is achieved for $\Omega = 0$. Increasing the spectral width, $\sigma$, reduces the range of frequencies $\Omega$ that are unstable without changing the maximum instability rate. Decreasing the initial power difference, $\Delta N^0$, reduces both the range of unstable modes $\Omega$ and the instability growth-rate. The influence of $\alpha$ is more subtle and reducing $\alpha$ can lead to a shift of the maximum instability rate from $\Omega = 0$ to $\Omega \neq 0$.
\subsubsection{Stability analysis of the normal correlators, Eqs.(\ref{eq:deltanx_vl_supp}-\ref{eq:deltany_vl_supp})}

Following a procedure similar to that for the anomalous correlator, we now carry out the stability analysis for the normal correlators whose evolution is governed by the linearized Eqs.(\ref{eq:deltanx_vl_supp}-\ref{eq:deltany_vl_supp}).
Defining $\delta {\hat n}_\mu(\Omega,\omega,\lambda_n)=\int_0^z ds \int dt \delta {n}_\mu(t,\omega,z) \exp(-i\Omega t -\lambda_n s)$, we obtain the set of coupled equations:
\begin{align}
\delta {\hat n}_x(\Omega,\omega,\lambda_n) &= 
\frac{i\gamma \Omega \partial_\omega n_x^0(\omega)}{2\pi(\lambda_n + 2 i \beta \omega \Omega)} 
\int  2\delta {\hat n}_x(\Omega,\omega_1,\lambda_n)+\kappa \delta {\hat n}_y(\Omega,\omega_1,\lambda_n) d\omega_1,\\
\delta {\hat n}_y(\Omega,\omega,\lambda_n) &= 
\frac{i\gamma \Omega \partial_\omega n_y^0(\omega)}{2\pi(\lambda_n + 2 i \beta \omega \Omega)} 
\int  2\delta {\hat n}_y(\Omega,\omega_1,\lambda_n)+\kappa \delta {\hat n}_x(\Omega,\omega_1,\lambda_n) d\omega_1,
\end{align}
The dispersion relation $\lambda_n(\Omega)$ for the normal correlator is obtained by solving the integral equation:
\begin{align}
&\big(1-2{\cal I}_x(\Omega,\lambda_n) \big) \big(1-2{\cal I}_y(\Omega,\lambda_n) \big)=\kappa^2 {\cal I}_x(\Omega,\lambda_n) {\cal I}_y(\Omega,\lambda_n),
\label{eq:lambda_n}
\end{align}
where the integrals ${\cal I}_\mu(\Omega,\lambda_n)$ are given by 
\begin{align}
{\cal I}_\mu(\Omega,\lambda_n)=\frac{i\gamma \Omega }{2\pi}\int \frac{\partial_\omega n_\mu^0(\omega)}{\lambda_n + 2 i \beta \omega \Omega} d\omega = -\frac{\gamma \beta \Omega^2 }{\pi}\int \frac{n_\mu^0(\omega)}{(\lambda_n + 2 i \beta \omega \Omega)^2} d\omega, \quad \mu=x,y.
\end{align}
We consider initial Gaussian spectra of the form: $n_\mu^0(\omega)=\frac{\sqrt{2\pi}}{\sigma} N_\mu^0 \exp(-\omega^2/(2 \sigma^2))$. We have solved numerically Eq.(\ref{eq:lambda_n}) to compute the dispersion relation $\lambda_n(\Omega)$. In the weakly nonlinear regime verifying $L_{lin} \sim (\beta \sigma^2)^{-1} \ll L_{nl} \sim (\gamma N)^{-1}$, the numerical analysis shows that the normal correlator is always stable, ${\rm Re}\big(\lambda_n(\Omega)\big) <0$.   

\subsection{Nonlinear dynamics with homogeneous statistics: Derivation of Eq.(8) (main text)} 

The analysis of the dispersion relation has revealed that, in general, the homogeneous mode $\Omega=0$ is the most unstable mode, see Fig.~\ref{fig:stability}. Here, we analyze the nonlinear dynamics of Eq.(\ref{eq:nx_vl_supp}-\ref{eq:m_vl_supp}) in the limit of stationary statistics:
\begin{align}
\partial_z n_x(\omega,z) =& +4 \gamma\kappa M^r m^i(\omega)  ,
\label{eq:nx_vl_stat} \\
\partial_z n_y(\omega,z) =& -4 \gamma\kappa M^r m^i(\omega)  ,
\label{eq:ny_vl_stat} \\
\partial_z m(\omega,z) =& 
-i  \big( \gamma(2-\kappa)(N_y-N_x)-2\alpha   \big) m 
- 2i\gamma\kappa \big( n_x - n_y \big) M^r .
\label{eq:m_vl_stat}
\end{align}
Note that, $\left< {\tilde u}_y(\omega_1,z) {\tilde u}_x^*(\omega_2,z) \right>=2\pi m(\omega_1,z) \delta(\omega_1-\omega_2)$ under the assumption of homogeneous statistics (${\tilde u}_\mu(\omega,z)=\int u_\mu(t,z) \exp(-i\omega t) dt$ being the Fourier transform of the field), so that anomalous correlations between $u_x$ and $u_y$ arise only at the same frequency.

By integration over $\omega$, we get the following system for the normal and anomalous correlators:
\begin{align}
\partial_z \Delta N &= -8 \gamma \kappa M^r M^i ,
\label{eq:DelaN_red}\\
\partial_z M^r &= \big( \gamma (2- \kappa)\Delta N -2\alpha \big) M^i , 
\label{eq:Mr_red}\\
\partial_z M^i &= - \big(\gamma (2- 3 \kappa)\Delta N -2\alpha \big) M^r .
\label{eq:Mi_red}
\end{align} 
where $\Delta N(z)=N_y(z)-N_x(z)$, and we recall that $N_\mu(z) = \frac{1}{2\pi}\int n_\mu(\omega,z) d\omega$, $M(z) = \frac{1}{2\pi}\int m(\omega,z) d\omega$. 
Introducing the Stokes vector ${\bm S}(z)=\big( \Delta N(z), 2 M^r(z), 2 M^i(z) \big)^T$, the dynamics (\ref{eq:DelaN_red}-\ref{eq:Mi_red}) can be recast as a rotation of this vector on the Poincar\'e sphere (see Eq.(8) in the main text):
\begin{align}
&\partial_z {\bm S}(\tau,z) = {\bm R}(z) \times  {\bm S}(z), 
\label{eq:S_dyn}\\
&{\bm R}(z) = \big( 2\alpha-\gamma (2-\kappa) S_1(z), \ -2\gamma \kappa S_2(z), \ 0  \big)^T.
\label{eq:S_R_dyn}
\end{align}
Because ${\bm S}.\partial_z {\bm S}=0$, the vector ${\bm S}(z)$ evolves on the surface of the Poincar\'e sphere of constant radius 
\begin{align}
S_0=\Big( \sum_{j=1}^3 S_j^2  \Big)^{1/2}=\sqrt{\Delta N(z)^2 + 4|M(z)|^2}= {\rm const}.
\label{eq:S_0}
\end{align}
Consequently, the degree of polarization is conserved during  evolution,  ${\cal P}(z)=S_0/N=$const.

Considering the case $\kappa=2/3$ relevant to our experiments: The fixed point $S_1=-S_0$ is stable. Conversely, as discussed in the main text through Eq.(8), the fixed point $S_1=+S_0$ becomes unstable for $\alpha < \alpha_c = \frac{2}{3}\gamma S_1$, giving rise to two new stable fixed points located at $S_1=3\alpha/(2\gamma), S_2=0, S_3=\pm \sqrt{S_0^2-S_1^{2}}$.  

%Note that the good agreement shown in Fig.~2(b) (main text) between the simulations of the AC-KE~(\ref{eq:S_dyn}-\ref{eq:S_R_dyn}) and the NLSE (\ref{eq:nls_x}-\ref{eq:nls_y}) is obtained for a significantly large value of $|E/U| \simeq 158$. As $|E/U|$ is reduced, higher-order (beyond quadratic) nonlinear terms in the kinetic equation become increasingly important and should be taken into account. The formulation of such a generalized kinetic theory constitutes an open problem, as discussed in the conclusion of the main text.

The Poincar\'e–Stokes formalism employed here differs fundamentally from the one commonly used to describe the polarization dynamics of fully coherent stationary waves \cite{agrawal}. First, the Stokes vector represents the  correlation functions of an incoherent random wave, rather than the coherent field amplitudes. Second, in the  coherent stationary case, the radius of the Poincar\'e sphere $S_0$, is determined solely by the total power $N$; consequently, all solutions with fixed power $N$ evolve on the surface of a single sphere. In contrast, in the present formulation the sphere radius also depends on the degree of polarization, $S_0 = {\cal P} N$. As a result, $S_0$ varies with the initial conditions even when the total power is held fixed ($N=$const): The full set of solutions is not described by a single sphere, but instead by an ensemble of spheres with different radii.

\section{Experimental set-up}

\begin{figure}[!h]
    \centering
    \includegraphics[width=0.7\linewidth]{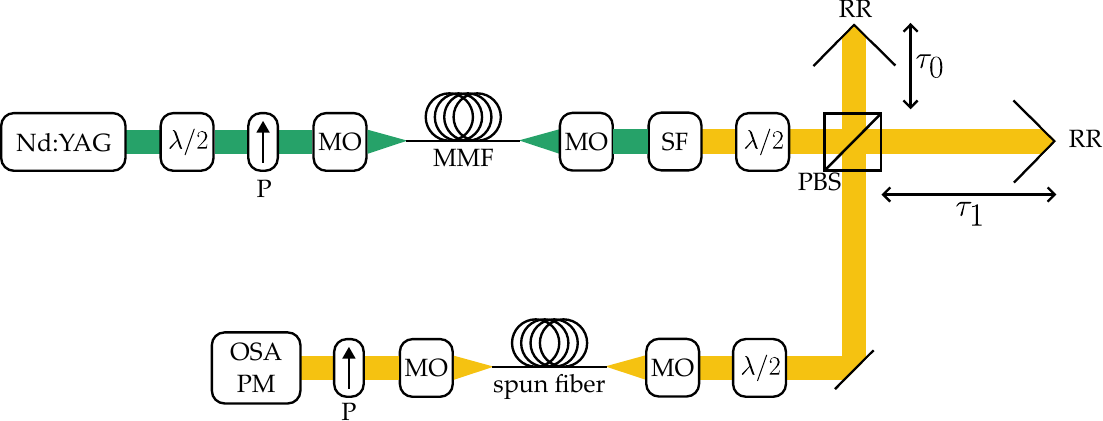}
    \caption{
    \baselineskip 10pt
    \textbf{Experimental set-up.} Laser source (Nd:YAG), halfwave plate ($\lambda/2$), polarizer (P), microscope objectives (MO), multimode fiber (MMF), spectral filter (SF), polarization beam splitter (PBS), retroreflector (RR), optical spectrum analyzer (OSA) polarimeter (PM).}
    \label{fig:set-up}
\end{figure}

The experimental set-up is presented in Fig.~\ref{fig:set-up}.

The source is a Nd:YAG laser delivering subnanosecond pulses (400ps) at $\lambda$ =532nm. Using a microscope objective ($\times20$), the beam is first injected into a standard GRIN multimode fiber \cite{krupa16}. 
During the nonlinear propagation, a temporally incoherent pulse is generated via the Raman beam clean-up effect which allows to produce Stokes beam with good spatial beam quality \cite{terry,pourbeyram}. At the output of the multimode fiber (MMF), we spectrally filter the second Raman Stokes, centered around $\lambda = 558\ \rm{nm}$ with FWHM of $2\ \rm{nm}$ and characterized by a nearly Gaussian beam shape. 
The beam is then passed through a half-wave plate to control the power on each polarization axis before a polarization beam splitter (PBS). The half-wave plate is oriented such that there is an equal amount of power on the two orthogonal polarization axes. Each polarization propagates for a different time $\tau_0$ and $\tau_1$ using free-space delay lines before being recombined through the PBS. We introduce a time delay, $\Delta\tau = \tau_1 - \tau_0 \approx 15\ \rm{ps}$ which is larger than the coherence time of the pulse ($<1\ \rm{ps}$), hence removing the mutual coherence between the two polarization axes, i.e. $|M| \approx 0$.
The beam is then passed through another half-wave plate in order to align the two polarizations with the principal axes of the weakly birefringent fiber. The weak birefringence is obtained by carefully winding a spun fiber onto a spool with a diameter of $14.5$ cm. Spun fibers are made by rapidly spinning a preform during the drawing process. This results in a fast rotation of the slow and fast axis of the fiber, leading the averaged net birefringence to be zero. The mechanical torsion from the winding introduces a weak, controlled linear birefringence $dn \approx 10^{-7}$. 
The axes of the weakly birefringent fiber were identified by measuring the polarization modulation instability of the coherent Nd:YAG beam directly injected \cite{millot_josab98}.
Microscope objectives ($\times 40$) were used to inject and collect light in and out of the fiber.
At the output of the fiber, we measure the optical spectrum and the power on each polarization axis. To characterize optical spectra, we used the OSA YOKOGAWA-AQ6374.
Furthermore, the temporal profile of the pulse was monitored before the injection into the weakly birefringent fiber using a high-speed oscilloscope combined with 12-GHz photodiode.
To measure the complex anomalous correlator, we used a homemade polarimeter made of a combination of quarter-wave, half-wave plates, polarizers, and powermeter as commonly applied in standard polarization measurement.

\section{Numerical simulations of the experimental regime}

%To interpret the experimental observations, we numerically integrated the coupled NLSE Eqs.~(\ref{eq:nls_x}-\ref{eq:nls_y}), with explicit physical parameters :
%\begin{eqnarray}
%    \frac{\partial u_x}{\partial z} &=& \beta_{1x}\frac{\partial u_x}{\partial t} + \frac{i\beta_{2}}{2}\frac{\partial^2 u_x}{\partial t^2} - i\gamma\left( |u_x|^2 + \frac{2}{3}|u_y|^2\right)u_x - \frac{i\gamma}{3}u_x^*u_y^2e^{-2i\Delta\beta z}, \\
%    \frac{\partial u_y}{\partial z} &=& \beta_{1y}\frac{\partial u_y}{\partial t} + \frac{i\beta_{2}}{2}\frac{\partial^2 u_y}{\partial t^2} -i\gamma\left( |u_y|^2 + \frac{2}{3}|u_x|^2\right)u_y + \frac{i\gamma}{3}u_y^*u_x^2e^{2i\Delta\beta z}.
%\end{eqnarray}
%In terms of the birefringence, $dn$, of the fiber, we have $\Delta\beta_1 = dn/c$ and $\Delta\beta = (2\pi/\lambda_0)dn$, with $c$ the speed of light and $\lambda_0$ the pulse central wavelength \blue{We should have: $\Delta \beta = \omega_0 dn/c=(2\pi/\lambda_0) dn$}\red{Oui, j'ai inversé $\Delta \beta$ et $\Delta \beta_1$. Mais on peut enlever les équation oui}\\
%\blue{Je pense qu'on peut supprimer l'ecriture redondante des equations:\\ 
To interpret the experimental observations, we have numerically integrated the coupled NLS Eqs.~(\ref{eq:nls_x}-\ref{eq:nls_y}), where  $\beta=\beta_2/2$, with $\beta_2$ the group-velocity dispersion coefficient \cite{agrawal}. The coherent coupling parameter $\alpha=\Delta \beta/2$ is related to the fiber birefringence $\Delta \beta=(2\pi/\lambda_0) dn$, with $dn=|n_x-n_y|$  the refractive index difference among the axes 
%$c$ the speed of light 
and $\lambda_0$ the pulse central wavelength. Furthermore, in the numerical simulations of the experiments we have included the minor correction provided by the group-velocity difference of the waves $u_x$ and $u_y$, which was found to have a negligible impact.
The fiber used in the experiment was the same as in Ref.\cite{millot_josab98,millot_ol98}, with parameters $dn\approx1\times10^{-7}$, effective area $A_{\rm{eff}} \approx 15\ \mathrm{\mu m^2}$ and $\beta_2 \approx 60 \ \mathrm{fs^2/mm}$ at $\lambda_0 \approx 558 \ \mathrm{nm}$, nonlinear refractive index $n_2 = 3.2 \times 10^{-20}$m$^2/$W, and corresponding nonlinear parameter $\gamma = 2 \pi n_2 /(\lambda_0 A_{\rm{eff}}) \approx 0.024\ \rm{W^{-1}m^{-1}}$.\\
The spectrum and temporal profile of the initial pulses were recorded experimentally before injection into the fiber and served as initial conditions in the numerical simulations (spectrum with a FWHM of $\approx 2$-nm and $100$-ps long pulses). A random phase was added on each spectral component to generate an incoherent temporal wave.

\subsection{Thermalization in the experimental regime}

\begin{figure}[!h]
    \centering
    \includegraphics[width=0.4\linewidth]{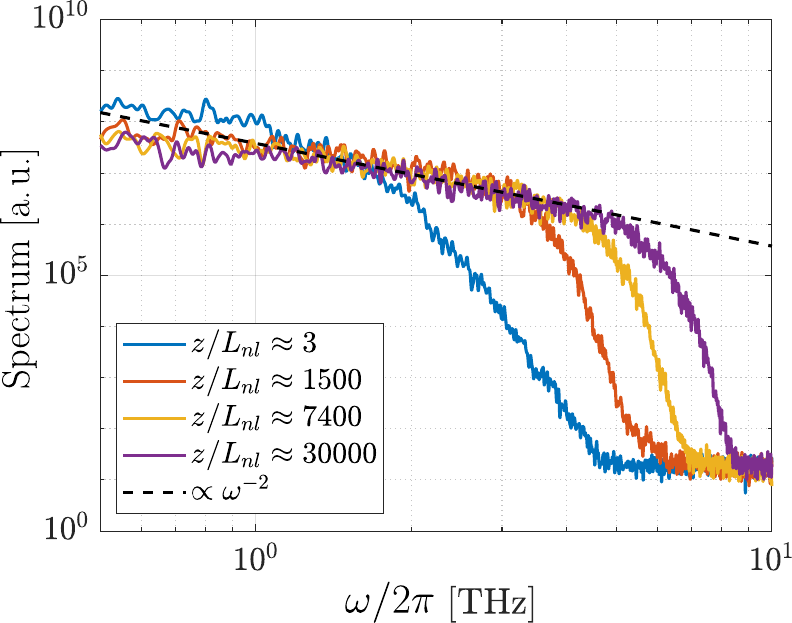}
    \caption{
    \baselineskip 10pt
    Local quasi-equilibrium in the experimental regime. Numerical simulation of the NLSE in the experimental configuration for a peak power of $\approx 18\ \rm{W}$. The spectral window in which the field is in local quasi-equilibrium increases as $z$ increases.}
    \label{fig:S2}
\end{figure}

In the main text, we have reported the repolarization effect. Due to experimental limitations, we have been able to investigate the early stage of the thermalization process. As discussed in \cite{zanaglia2025spatio}, the optical field approaches at each $z$ a local quasi-equilibrium within a limited spectral window. As $z$ increases, this spectral window grows and the spectral distribution follows, locally, an equilibrium state, characterized by the power-law decay $\sim \omega^{-2}$ in the spectral tail. This is illustrated in Fig.~\ref{fig:S2}, which shows the evolution of the spectrum as $z$ increases. We can see that the spectral window of local equilibrium increases as the field propagates over large propagation lengths, exhibiting the thermalization process in the experimental configuration. This confirms that the experimental results reported through Fig.~3 (main text) provide a signature of the thermalization process. 

%\end{widetext}

\end{document}